\documentclass{aa}  

\usepackage{natbib}
\usepackage[FIGTOPCAP]{subfigure}
\usepackage{multirow}
\usepackage{txfonts}
\usepackage[dvipsnames]{xcolor}
\usepackage{tablefootnote}
\usepackage{scrextend}
\usepackage[final]{pdfpages}

\usepackage{hyperref}
\hypersetup{
    colorlinks   = true,
    citecolor=RoyalBlue,
    filecolor=black,
    linkcolor=PineGreen,
    urlcolor=RoyalBlue
}

\begin{document}

\defcitealias{1997ApJ...489..698H}{H97}

   \title{Massive pre-main sequence stars in M17\thanks{Based on observations collected at the European Southern Observatory at Paranal, Chile (ESO programmes 60.A-9404(A), 085.D-0741, 089.C-0874(A), and 091.C-0934(B)).}
   }

   \author{M.C. Ram\'irez-Tannus \inst{1}
          \and
          L. Kaper\inst{1}
          \and 
          A. de Koter\inst{1,2}
           \and
          F. Tramper\inst{3}
          \and
          A.Bik\inst{4}
          \and
          L.E. Ellerbroek\inst{1}
          \and 
          B.B. Ochsendorf\inst{5}
          \and 
          O.H.~Ram\'irez-Agudelo\inst{6}
          \and
          H. Sana\inst{2}                 
}

   \institute{Anton Pannekoek Institute for Astronomy, University of Amsterdam,
              Science Park 904, 1098 XH Amsterdam, The Netherlands;\\
              \email{m.c.ramireztannus@uva.nl}
		\and 
		Institute of Astrophysics, 
		Universiteit Leuven, Celestijnenlaan 200 D, 3001 Leuven, Belgium
		\and
		 European Space Astronomy Centre (ESA/ESAC), Operations Department, 
		 Villanueva de la Ca\~{n}ada  (Madrid), Spain 
		 \and
		Department of Astronomy, Stockholm University, 
		Oskar Klein Center, SE-106 91 Stockholm, Sweden
		\and
		Department of Physics and Astronomy, 
		Johns Hopkins University, Baltimore, MD, USA
		\and
		UK Astronomy Technology Centre, Royal Observatory Edinburgh, 
		Blackford Hill, Edinburgh, EH9 3HJ, UK \\
}

   \date{\today}
 
   \abstract {The formation process of massive stars is still poorly understood. Massive young stellar objects (mYSOs) are deeply embedded in their parental clouds, they are rare and thus typically distant, and their reddened spectra usually preclude the determination of their photospheric parameters. M17 is one of the best studied \ion{H}{ii} regions in the sky, is relatively nearby, and hosts a young stellar population. With X-shooter on the ESO {\it Very Large Telescope} we have obtained optical to near-infrared spectra of candidate mYSOs, identified by \citet{1997ApJ...489..698H}, and a few OB stars in this region. The large wavelength coverage enables a detailed spectroscopic analysis of their photospheres and circumstellar disks. We confirm the pre-main sequence (PMS) nature of six of the stars and characterise the O stars. The PMS stars have radii consistent with being contracting towards the main sequence and are surrounded by a remnant accretion disk. The observed infrared excess and the (double-peaked) emission lines provide the opportunity to measure structured velocity profiles in the disks. We compare the observed properties of this unique sample of young massive stars with evolutionary tracks of massive protostars by \citet{2009ApJ...691..823H}, and propose that these mYSOs near the western edge of the \ion{H}{ii} region are on their way to become main-sequence stars ($\sim 6 - 20$~$M_{\odot}$) after having undergone high mass-accretion rates (${\dot{M}_{\rm acc}} \sim 10^{-4} - 10^{-3}$~$M_{\odot}$~$\rm yr^{-1}$). Their spin distribution upon arrival at the zero age main sequence (ZAMS) is consistent with that observed for young B stars, assuming conservation of angular momentum and homologous contraction.}


   \maketitle
%
%________________________________________________________________
                                                 
\section{Introduction}
\label{Introduction}

In the past decades significant progress has been made regarding the understanding of star formation. However, the formation of massive stars -- the only mode of star formation observable in external galaxies -- remains a key unsolved problem \citep{2007ARA&A..45..481Z, 2007prpl.conf..165B, 2014prpl.conf..149T}. One of the main reasons is that observations of the earliest phases in the life of a massive star are scarce. This is explained by the expected short formation timescale ($\sim 10^{4} - 10^{5}$~yr) of massive stars, and the severe extinction ($A_{V} \sim 10 - 100$~mag) by which the surrounding gas and dust obscure their birth places from view. Additionally, massive stars are rare and, as a consequence, located at relatively large distances.

Observational and theoretical evidence is accumulating that the formation process of massive stars is through disk accretion, similar to low-mass stars. This despite the physical processes acting near hot, massive stars (e.g., radiation pressure, ionisation) that counteract the accretion flow onto the forming star \citep[e.g.,][]{1987ApJ...319..850W, 2006ApJ...637..850K,  2009Sci...323..754K, 2016ApJ...823...28K}. In the accretion scenario, one expects that soon after the gravitational collapse of the parental cloud the pre-stellar core is surrounded by an extended accretion disk \citep[size $\sim 1000$~AU; e.g.,][]{2004ApJ...601L.187B,  2005A&A...434.1039C, 2010ApJ...715..919S, 2016A&ARv..24....6B}. Apparently this is not a very efficient process, as most of these high-mass protostars seem to drive outflows \citep[e.g.,][]{2001ApJ...552L.167Z}.  Already before arriving on the main sequence, the young massive star is expected to produce a strong ultraviolet (UV) radiation field ionising its surroundings \citep{1996A&A...307..829B}. At this stage the object will become detectable at radio and infrared wavelengths through the heated dust and recombination of hydrogen in an expanding hyper- or ultra-compact \ion{H}{ii} (UCH{\sc ii}) region \citep[e.g.,][]{2002ASPC..267....3C}. Unfortunately, only very little is known about the physical properties of the central (massive) stars at this stage of formation \citep[e.g.][]{2003A&A...405..175M}.

Given the short formation time of massive stars, the accretion rate must be high \citep[up to $10^{-3}$~$M_{\odot}$~yr$^{-1}$,][]{2010ApJ...721..478H}. At these accretion rates, a massive young stellar object (mYSO) is expected to bloat up to about 100~$R_{\odot}$, resulting in a relatively low effective temperature and modest UV luminosity. The accretion process is unlikely to be constant with time; the blob structure observed in Herbig Haro outflows indicates that strong variations occur in the mass in- and outflow rate of young intermediate-mass stars \citep[e.g.][]{2001ARA&A..39..403R, 2013A&A...551A...5E}. Simulations of different accretion models applied to mYSOs show that the accretion rate increases as the mYSO grows in mass \citep{2011MNRAS.416..972D}. Once the accretion rate diminishes, the ``bloated'' pre-main-sequence (PMS) star contracts to the main sequence on the Kelvin-Helmholtz timescale. Recently, candidates for such bloated, massive PMS stars have been spectroscopically confirmed \citep{2011A&A...536L...1O, 2012ApJ...744...87B, 2013A&A...558A.102E, 2015A&A...578A..82C}.

The environment (multiplicity, clustering) and the large distances to massive (proto)stars make their circumstellar disks very difficult to resolve \citep{2016A&ARv..24....6B}. In contrast, disks around intermediate-mass stars (2 < $M_{\star}$ < 8~$M_{\odot}$) have been characterised using (sub)millimeter as well as optical and near-infrared spectroscopy \citep[see][]{2004A&A...427L..13B, 2010MNRAS.408.1840W, 2011ApJ...732L...9E, 2014A&A...561A...2A}. The rotational structure of the disk provides information about the role of magnetic fields that can slow down the rotation rate below pure Keplerian. Super-Keplerian rotation of the disk could indicate that the inner disk contributes significantly to the gravitational potential of the system \citep{2016A&ARv..24....6B}. In the mass range from 20--30~$M_{\odot}$ circumstellar disks have been detected, but there is no consensus about their general properties.

M17, located in the Carina-Sagittarius spiral arm of the Galaxy is one of the best studied giant \ion{H}{ii} regions \citep[$L=3.6\times10^{6}L_{\odot}$,][]{2007ApJ...660..346P}. Its distance has been accurately determined by measuring the parallax of the $\rm CH_{3}OH$ maser source G15.03-0.68: $d = 1.98^{+0.14}_{-0.12}$~kpc \citep{2011ApJ...733...25X}. The bright blister \ion{H}{ii} region is embedded in a giant molecular cloud complex, and divides the molecular cloud into two components: M17 South and M17 North, containing a total gas mass (molecular and atomic) of about $6 \times 10^{4}$~$M_{\odot}$ \citep{2009ApJ...696.1278P}. The centre of the \ion{H}{ii} region hosts the cluster NGC~6618 including 16 O stars and over 100 B stars \citep{1980A&A....91..186C, 2008ApJ...686..310H}, providing the ionising power for the \ion{H}{ii} region. Many of the OB stars are suspected to be binaries \citep{2008ApJ...686..310H, 2009ApJ...696.1278P}, which would explain why they are more luminous than expected from their spectral type. NGC~6618 is a young cluster, its age is estimated at about 1~Myr \citep{1997ApJ...489..698H, 2007ApJS..169..353B, 2009ApJ...696.1278P}, while the surrounding molecular cloud hosts pre-main-sequence stars. The photo-dissociation region to the southwest of NGC~6618 includes several candidate  mYSOs: the hyper-compact \ion{H}{ii} region M17-UC1 surrounded by a circumstellar disk \citep{2007ApJ...656L..81N}, and IRS5, a young possibly quadruple system of which the primary star, IRS5A (B3--B7 V/III), is a high-mass protostellar object \citep[$\sim$9~$M_{\odot}$, $\sim 10^{5}$~yr, ][]{2015A&A...578A..82C}. \citet{1997ApJ...489..698H}  \citepalias[hereafter ][]{1997ApJ...489..698H} discovered a sample of high-mass (5--20~$M_{\odot}$) young stellar objects in the same area. Their SEDs display a near-IR excess, some show CO bandhead and (double-peaked) Pa~$\delta$ emission, likely due to the presence of a circumstellar disk. These objects are the subject of study in this paper.

We are now on the verge of being able to quantitatively test the predictions of massive star formation models. Several samples of mYSOs have been presented \citep[e.g.,][]{2006A&A...455..561B, 2011MNRAS.410.1237U, 2016ApJ...832...43O}. However, so far only for a handful of mYSOs the photospheric spectrum has been detected. We have obtained high-quality spectra with X-shooter on the ESO {\it Very Large Telescope} (VLT) of candidate mYSOs, first identified by \citetalias{1997ApJ...489..698H}, and a few other O and B stars  in M17. The large wavelength coverage of the X-shooter spectra (and additional infrared photometry) allows for a detailed analysis of the spectral energy distribution, including an infrared excess produced by the dust component in the accretion disk. The observed (double-peaked) emission lines provide the opportunity to study the dynamical structure of the gaseous component of the disk.

This paper is organised as follows. In Sect.~\ref{P1:VLT/X-shooter spectroscopy} we present our target sample, observations and data reduction procedure. In Sect.~\ref{P1:Spectral classification} we provide the spectral classification of the (pre-) main-sequence OB stars in our sample. After that (Sect.~\ref{P1:SEDs}) we present their spectral energy distributions. To further constrain the properties of the mYSOs we modeled the optical spectra using \texttt{FASTWIND}. These results are shown in Sect.~\ref{P1:FASTWIND}, where we also place the PMS objects in the Hertzsprung-Russell diagram and compare their position with PMS tracks and isochrones. Additional evidence for the presence of circumstellar disks is obtained from near-infrared spectroscopy (Sect.~\ref{P1:NIR}). We discuss the age distribution of the massive PMS population in M17, the extinction towards this region, the presence of circumstellar disks, and the spin properties of the young massive stars in Sect.~\ref{P1:Discussion}. We summarise our conclusions in Sect.~\ref{P1:Conclusions}.

%__________________________________________________________________

\section{VLT/X-shooter spectroscopy}
\label{P1:VLT/X-shooter spectroscopy}

We obtained optical to near-infrared (300--2500~nm) spectra of candidate mYSOs, and some OB~stars in M17. We used the X-shooter spectrograph mounted on UT2 of the VLT \citep{2011A26A...536A.105V}. The spectra of B275 and B358 were obtained as part of the X-shooter science verification program; B275 has been subject of study in \citet{2011A&A...536L...1O}. For some objects multiple spectra have been taken. A log of the observations is listed in Table~\ref{P1:tab:xshobs}.

\begin{table*}:
 \centering 
 \caption{\normalsize{Journal of X-shooter observations.} The first column lists the object's number in the Master's thesis of \citet{1992MsT..........2B}; the second column provides alternative IDs, the CEN \citep{1980A&A....91..186C}, and OI \citep{1976PASJ...28...35O} numbers. The third and fourth columns list the right ascension $\alpha$ and declination $\delta$ (J2000). The fifth and sixth columns display the $V$- (see footnote for references) and $K$-band magnitude \citep[2MASS, ][]{2003yCat.2246....0C}; the seventh column shows the date of the observations. The exposure times for each arm are shown in columns 8 to 10. 
B111, B163, B164, and B253 are associated with NGC~6618 \citep{2009ApJ...696.1278P}.}             
\begin{minipage}{\textwidth}.
 \centering 
\renewcommand{\arraystretch}{1.4}
\setlength{\tabcolsep}{5pt}
\begin{tabular}{llcccccccc}
\hline
\hline
\multicolumn{2}{c}{Object} & $\alpha$ (J2000) & $\delta$ (J2000)& $V$ & $K$ & Date & \multicolumn{3}{c}{Exp. time [s]} \\
M17-	&	CEN (OI)	&	(h m s)	&	($\degr$  $\arcmin$  $\arcsec$)	&	mag	&	mag	&		&	UVB	&	VIS	&	NIR	\\
\hline																					
B111	&	2 (337)	&	18:20:34.67	&	-16:10:10.50	&	11.207\footnote{\label{P1:foot:AAVSO}AAVSO Photometric all sky survey (APASS) catalog: \url{https://www.aavso.org/apass}}	&	7.475	&	2013-07-17	&	2x60	&	2x90	&	2x10	\\
B163	&	-	&	18:20:30.95	&	-16:10:39.40	&	$-$	&	9.686	&	2010-09-17	&	2x300	&	2x300	&	2x300	\\
B164	&	25	&	18:20:30.90	&	-16:10.08.00	&	15.410\footnote{\label{P1:foot:Chini80}\citet{1980A&A....91..186C}}	&	8.758	&	2013-07-17	&	2x470	&	2x500	&	2x30	\\
B215 (IRS15)	&	-	&	18:20:28.70	&	-16:12:12.00	&	16.100\footnote{\label{P1:foot:Hoff08}\citet{2008ApJ...686..310H}}	&	10.004	&	2012-07-07	&	2x900	&	2x400	&	2x50	\\
B243	&	51	&	18:20:26.64	&	-16:10:03.70	&	17.800\footref{P1:foot:Chini80}	&	9.544	&	2012-07-06	&	2x900	&	2x870  	&	2x50 	\\
	&		&		&		&		&		&	2013-07-17	&	2x870	&	2x450	&	2x50	\\
B253	&	26	&	18:20:26.10	&	-16:11:18.00	&	15.740\footref{P1:foot:Chini80}	&	10.308	&	2013-07-17	&	2x470	&	2x500	&	2x30	\\
B268	&	49	&	18:20:25.35	&	-16:10:19.20	&	17.100\footref{P1:foot:Chini80}	&	9.494	&	2012-07-06	&	2x870	&	2x900  	&	2x50  	\\
	&		&		&		&		&		&	2012-07-06	&	2x870	&	2x300	&	2x50	\\
	&		&		&		&		&		&	2013-07-17	&	2x870	&	2x450	&	2x50	\\
B275	&	24	&	18:20:25.13	&	-16:10:24.56	&	15.550\footref{P1:foot:Chini80}	&	7.947	&	2009-08-11	&	4x685	&	8x285	&	12x11	\\
B289	&	31	&	18:20:24.39	&	-16:08:43.46	&	15.550\footref{P1:foot:Chini80}	&	9.178	&	2010-09-17	&	2x300	&	2x300	&	2x300	\\
	&		&		&		&		&		&	2012-07-06	&	2x870	&	2x300	&	2x50	\\
B311	&	16 (258)	&	18:20:22.70	&	-16:08:34.30	&	13.690\footref{P1:foot:Chini80}	&	8.884	&	2013-07-17	&	2x270	&	2x300	&	2x30	\\
B331	&	92	&	18:20:21.71	&	-16:11:18.40	&	20.100\footref{P1:foot:Hoff08}	&	8.946	&	2012-07-07	&	2x870	&	2x900	&	2x50	\\
B337	&	93	&	18:20:21.38	&	-16:10:41.20	&	$-$	&	9.343	&	2013-07-16	&	2x870	&	2x450	&	2x50	\\
B358	&	34	&	18:20:21.36	&	-16:09:59.60	&	$-$	&	7.788	&	2009-08-12	&	4x685	&	8x285	&	12x11	\\
\hline
\vspace{-15pt}
\end{tabular}
\renewcommand{\footnoterule}{}
\end{minipage}
\label{P1:tab:xshobs}
\end{table*}

  \begin{figure*}[ht]
   \centering
   \includegraphics[width=\hsize]{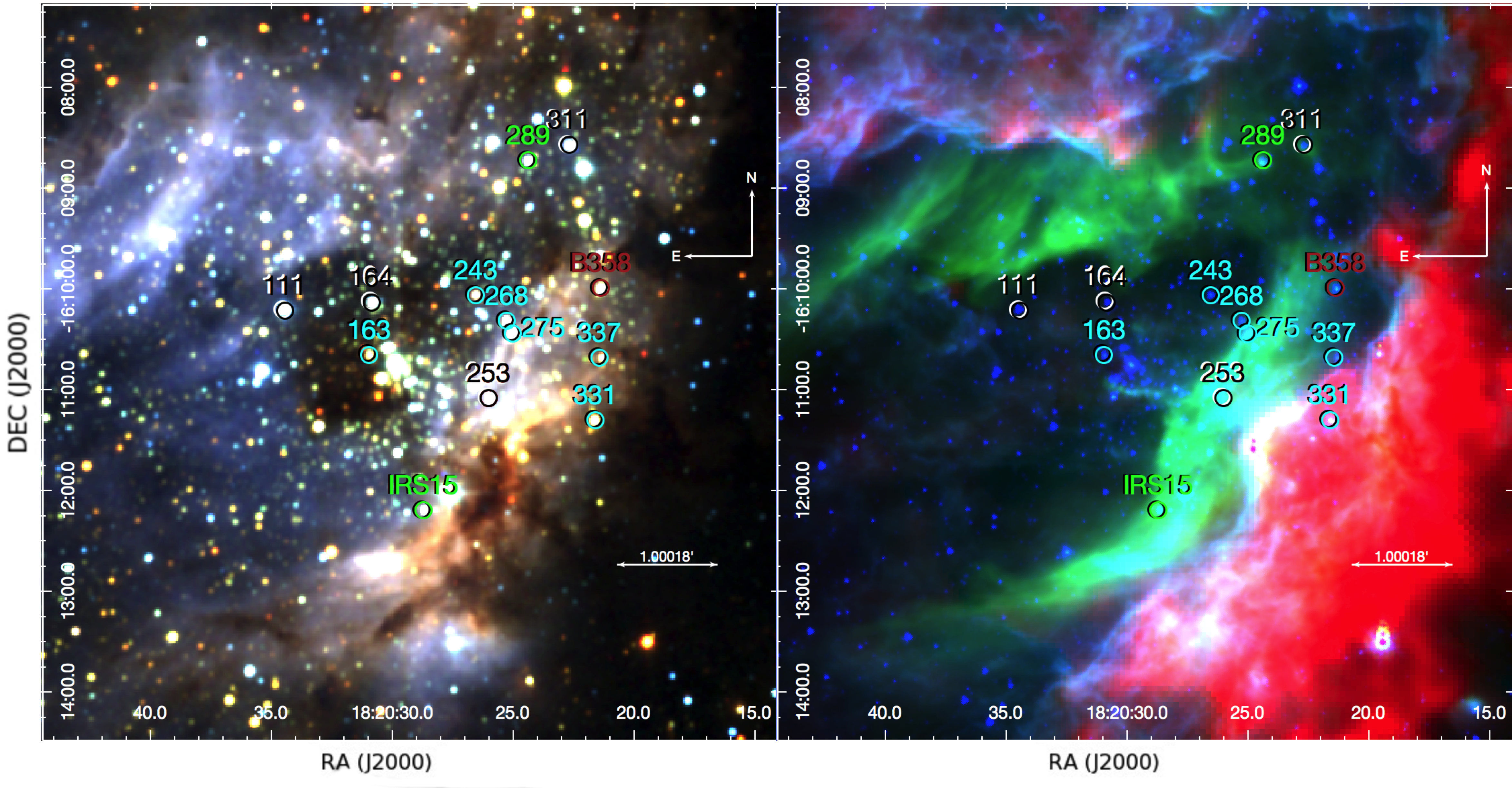}
      \caption{\emph{Left:} Near-infrared colour composite image of M17 based on 2MASS data: $J$ (blue), $H$ (green) and $K$ (red). Our X-shooter targets are labeled: in black we show the objects that have no indication of a disk, in green the objects that show NIR excess only longward of 3~$\mu$m in the photometry (points to a dusty disk), in blue the objects with emission lines in the X-shooter spectrum and NIR excess (identifying them as mYSOs), and in red the post-AGB star B358; B311 is associated with the bow shock M17-S1 \citep{2008ApJ...689..242P}. The central cluster NGC 6618 includes several dozen OB-type stars that ionise the surrounding region. At the western edge of the \ion{H}{ii} region several massive YSOs have been identified. Apparently, massive star formation is continuing in the irradiated molecular cloud. \emph{Right:} Mid-infrared colour composite image of M17 based on {\it Spitzer} and {\it Herschel} data. Blue: IRAC1 - 3.6~$\mu$m, green: IRAC4 - 8.0~$\mu$m, and red: PACS - 70~$\mu$m. Here the distribution of the heated dust (blue) becomes apparent, as well as PAH emission within the \ion{H}{ii} region (green).}
         \label{P1:fig:M17rgb}
   \end{figure*}
   
\subsection{Target sample}
\label{P1:Target sample}

We selected our targets based on the study of \citetalias{1997ApJ...489..698H} who obtained $K$-band spectra of the young massive star population of M17, complemented by optical and 1~$\mu$m spectroscopy for some of the moderately reddened objects. Several of the targets have also been studied by \citet{2008ApJ...686..310H} and some are included in the studies of \citet{2001A&A...377..273N}, \citet{2007ApJS..169..353B}, and \citet{2009ApJ...696.1278P}. \citetalias{1997ApJ...489..698H} reported that B111, B164, B253, and B311 do not present any NIR excess nor emission lines in their spectra and they are likely OB-star members of the central cluster NGC~6618. It is likely that all our targets belong to NGC~6681, at least the (massive) main sequence stars. 

Based on the NIR excess derived from their position in the NIR colour-colour diagram (CCD, see also Sect.~\ref{P1:CMD}) and displayed by the spectral energy distribution (SED, see also Sect.~\ref{P1:SEDs}), and the apparent presence of CO bandhead and/or double-peaked emission lines, the candidate mYSOs according to \citetalias{1997ApJ...489..698H} are B163, B243, B268, B275, B289, B331, and B337. We also observed B215 (IRS15): this mYSO candidate has been studied by \citet{2001A&A...377..273N}. The spectral types from \citetalias{1997ApJ...489..698H} and \citet{2008ApJ...686..310H} are listed in Table~\ref{P1:tab:Stellar_parameters}. We obtained an X-shooter spectrum of B358, but according to \citet{2013A&A...557A..51C} and \citetalias{1997ApJ...489..698H} the spectrum is that of an early-/mid-G supergiant. It has not been detected in X-rays \citep{2007ApJS..169..353B} and it is likely a background post-AGB star. From examining the X-shooter spectrum we support this conclusion.

Figure~\ref{P1:fig:M17rgb} (left panel) shows a colour composite image of M17 based on 2MASS $J$ (blue), $H$ (green) and $K$ (red) observations \citep{2006AJ....131.1163S}. The stars for which we collected X-shooter spectra are labeled. The area includes the bright nebular emission produced by the \ion{H}{ii} region and the central cluster NGC~6618. The mYSO candidates are predominantly located near the excited rim of the surrounding molecular cloud. This observation suggests that those objects could be the product of a later phase in the star formation process. In the right panel a colour composite is shown based on images taken with {\it Spitzer} \citep[GLIMPSE survey; IRAC1 - 3.6~$\mu$m, green: IRAC4 - 8.0~$\mu$m, ][]{2003PASP..115..953B} and {\it Herschel} \citep[red: PACS - 70~$\mu$m, ][]{2010A&A...518L...1P, 2010A&A...518L...2P}. This image provides insight into the (dusty) surroundings of the \ion{H}{ii} region. The 3.6~$\mu$m image traces the illuminated edges of the cloud irradiated by the massive stars, whereas the 8~$\mu$m emission is usually associated with emission features produced by polycyclic aromatic hydrocarbons (PAHs) and/or warm (100-200~K) dust that emits at these wavelengths when heated by ultraviolet radiation \citep{2009ApJ...696.1278P}. 

Throughout this paper we have color-coded the labels to present our results: the objects whose PMS nature is confirmed are plotted in blue, the OB stars in grey (black in figure~\ref{P1:fig:M17rgb}), the objects for which we cannot confirm the PMS nature in green and the post-AGB star in red.

\subsection{Colour-colour and colour-magnitude diagram of X-shooter targets in M17}
\label{P1:CMD}

\begin{figure}
   \centering
   \includegraphics[width=\hsize]{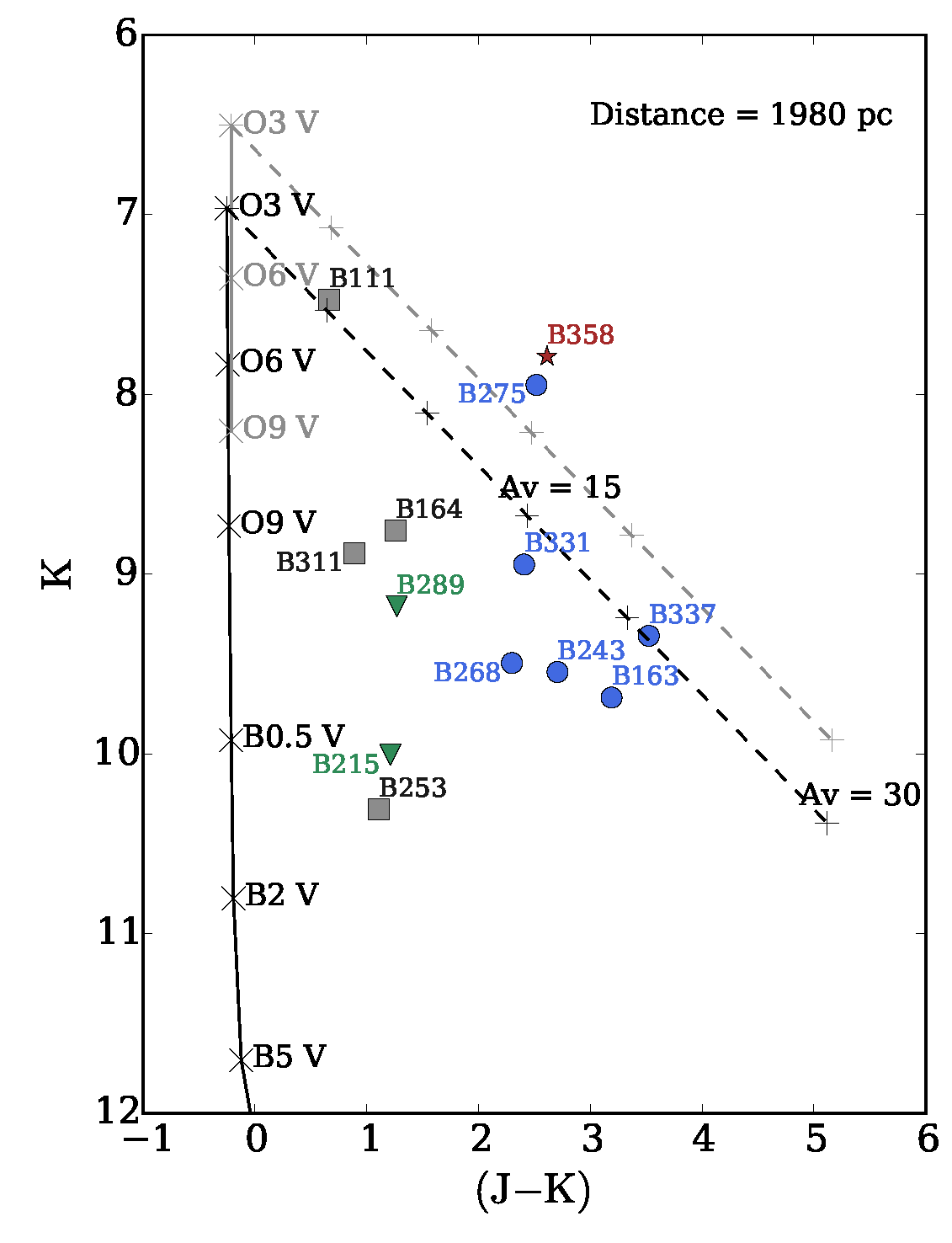}
      \caption{Near-infrared colour-magnitude diagram for our targets in M17. The objects for which we confirm their mYSO nature are marked by blue circles, the stars with NIR excess only longwards of 2.5~$\mu$m, but without emission lines, with green triangles, and the O and B main-sequence stars with grey squares. The red star corresponds to B358, most probably a background post-AGB star. The location of the zero-age main sequence is shown for a distance of 1.98~kpc with a solid black line \citep[\citetalias{1997ApJ...489..698H};][]{2000asqu.book.....C}. In grey we over-plot the updated parameters of O~V stars according to \citet{2006A&A...457..637M}.} The dashed line represents the reddening line for an O3~V star with $R_V=3.1$. Some of the mYSO candidates are located above the reddening line indicating the presence of a NIR excess.
         \label{P1:fig:CMD_JK}
   \end{figure}

The ($K$, $J-K$) colour-magnitude diagram is shown in Fig~\ref{P1:fig:CMD_JK}; the $J$, $H$, and $K$-band magnitudes are from the 2MASS database \citep{2003yCat.2246....0C}. Similar diagrams have been presented by \citetalias{1997ApJ...489..698H} and \citet{2008ApJ...686..310H}, but not using the recent distance estimate of 1.98~kpc \citep{2011ApJ...733...25X}. The zero-age main sequence (ZAMS) is marked with a drawn line; the dashed line represents the reddening vector. The ZAMS properties for spectral types later than B9 were taken from \citet{2000asqu.book.....C} and for the earlier type stars from \citetalias{1997ApJ...489..698H}. We over-plot the properties for the O dwarf stars from \citet{2006A&A...457..637M} for reference. We used the \citet{1989ApJ...345..245C} extinction law in order to plot the reddening lines. The candidate mYSOs have a large $(J-K)$ colour and a bright $K$-band magnitude. 

\begin{figure}
   \centering
   \includegraphics[width=\hsize]{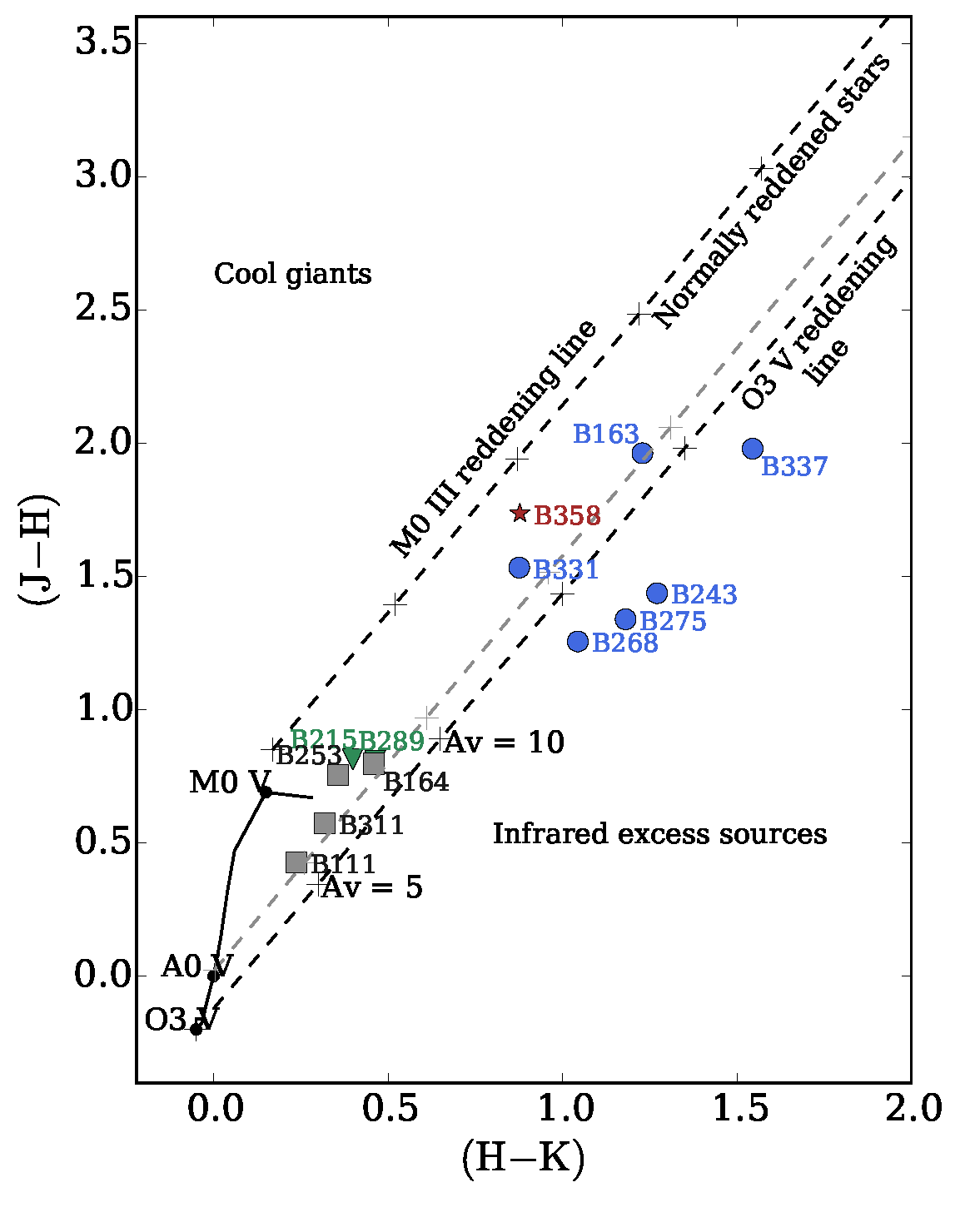}
      \caption{Near-infrared colour-colour diagram for our targets in M17. The colours and symbols are the same as in Fig.~\ref{P1:fig:CMD_JK}. The dashed lines represent the reddening lines for an O3~V and an M0~III star. For this we took the average Galactic value of $R_V=3.1$. In Sect.~\ref{P1:SEDs} we show that $R_V$ varies from 3.3 to 4.6. The solid line indicates the unreddened main sequence. Most confirmed mYSOs show evidence for the presence of a near-infrared excess.}
         \label{P1:fig:CCD_HK_JH}
   \end{figure}

Figure~\ref{P1:fig:CCD_HK_JH} presents the near-infrared colour-colour diagram including the OB-type stars for which an X-shooter spectrum has been obtained. In the colour-colour diagram a near-infrared excess becomes very apparent, and thus the possible presence of circumstellar material. The solid line represents the unreddened main sequence. The dashed lines indicate the reddening lines for an O3 dwarf and an M0 giant. Most of the mYSO candidates are located to the right of the reddening line of an O3~V star, consistent with the expectation that these objects host a circumstellar disk. B163, B289, and B331 are candidate mYSOs according to \citetalias{1997ApJ...489..698H}; however, they do not exhibit a near-infrared excess in the colour-colour diagram. Nevertheless, B331 and B163 are mYSO candidates according to our criteria: they present double-peaked emission lines, CO bandhead emission and/or infrared excess beyond the $K$-band. We will discuss the spectral energy distribution and the possible presence of an infrared excess in Sect.~\ref{P1:SEDs}, also at mid-infrared wavelengths. 

 \subsection{Reduction VLT/X-shooter spectra}

The X-shooter spectra were obtained under good weather conditions, 
with seeing ranging from 0\farcs{5} and 1\arcsec{} and clear sky. With the exception of the 2012 B289 spectrum and the 2009 B275 science verification spectrum, the spectrograph slit widths used were 1\arcsec{} (UVB, 300 -- 590 nm), 0\farcs{9} (VIS, 550 -- 1020 nm), and 0\farcs{4} (NIR, 1000 -- 2480 nm), resulting in a spectral resolving power of 5100, 8800, and 11300, respectively. The slit widths for the 2010 B275 observations were 1\farcs{6}, 0\farcs{9}, and 0\farcs{9} resulting in a resolving power of 3300, 8800, and 5600, respectively. For the 2012 B289 observations we used the 0\farcs{8}, 0\farcs{7}, and 0\farcs{4} slits corresponding to a resolving power of 6200, 11000, and 11300 for the UVB, VIS and NIR arms, respectively. The spectra were taken in nodding mode and reduced using the X-shooter pipeline \citep{2010SPIE.7737E..28M} version 2.7.1 running under the ESO Reflex environment \citep{2013A&A...559A..96F} version 2.8.4. 

\begin{figure*}
   \centering
   \includegraphics[width=\textwidth]{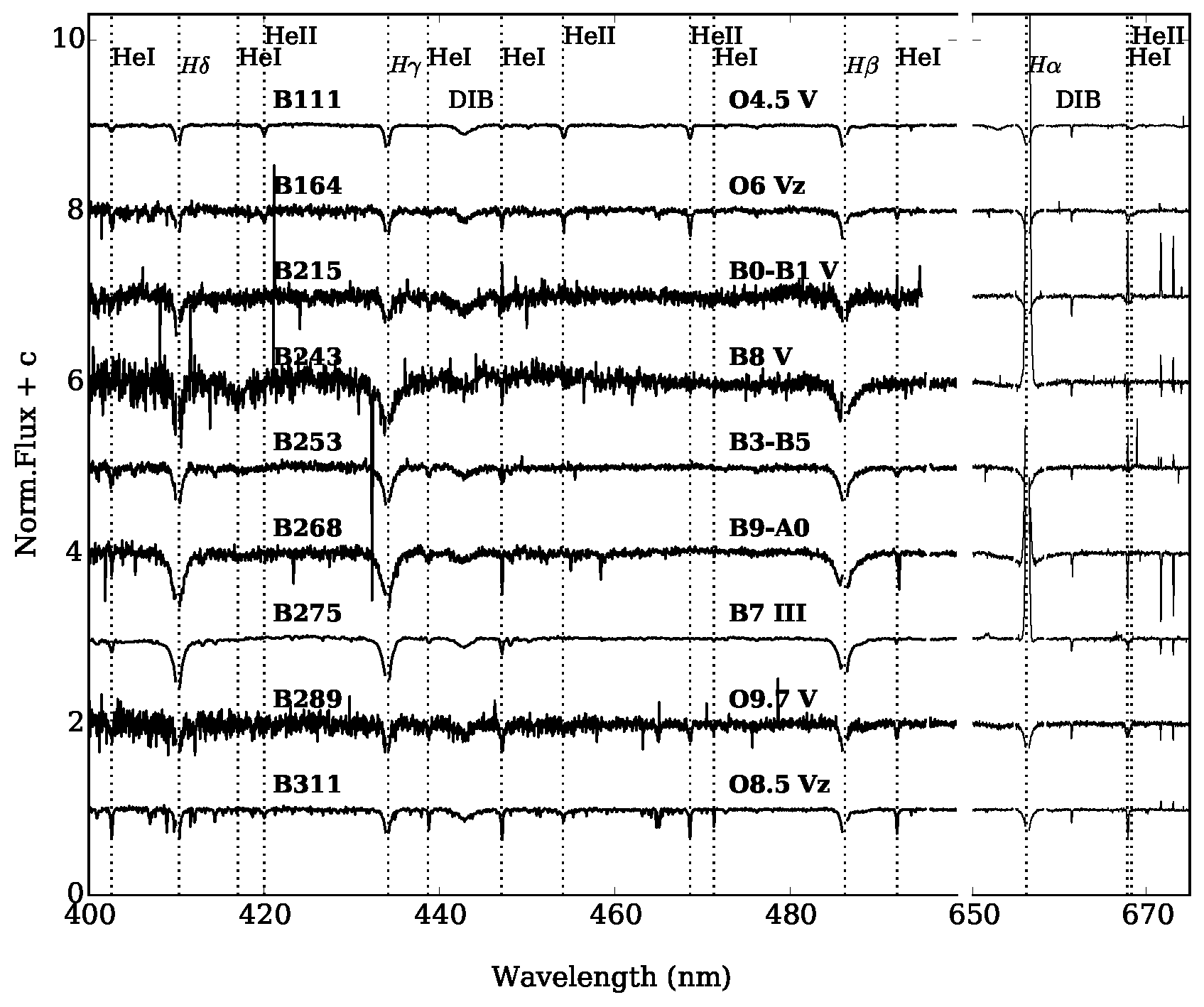}
      \caption{Blue region traditionally used for OB-star classification ({\it left}) and red region ({\it right}) including H$\alpha$ for the targets observed with X-shooter. The rest wavelengths of the fitted spectral lines for each star are indicated by the vertical dashed lines. The spectra include prominent diffuse interstellar bands (DIBs). The spectra have been clipped and smoothed to remove the residuals of the sky subtraction. B163, B331, and B337 are not included because they are severely extincted in this wavelength range.}
         \label{P1:fig:VisualSpec}
   \end{figure*}

\begin{figure*}
   \centering
   \includegraphics[width=\textwidth]{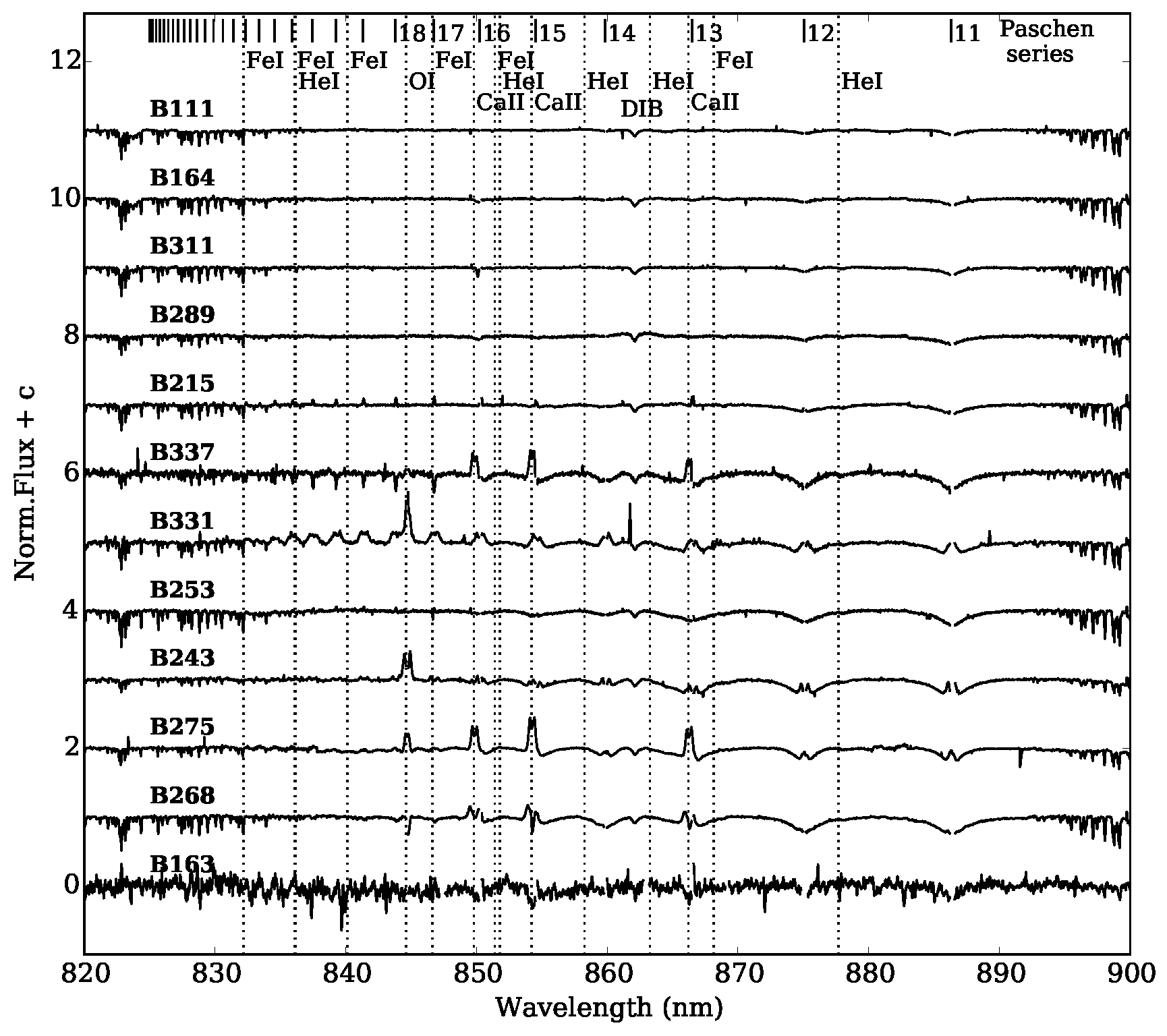}
      \caption{Calcium triplet region for each of the targets. In B337, B275 and B268 the \ion{Ca}{ii} triplet is clearly present; all targets exhibit Paschen absorption lines, some with a central emission component. In B331, B243 and B275 the \ion{O}{i} 8450 line is detected. The narrow features are either telluric or residuals from the subtraction of sky emission lines.}
         \label{P1:fig:CaT}
   \end{figure*}

\begin{figure*}
   \centering
   \includegraphics[width=\textwidth]{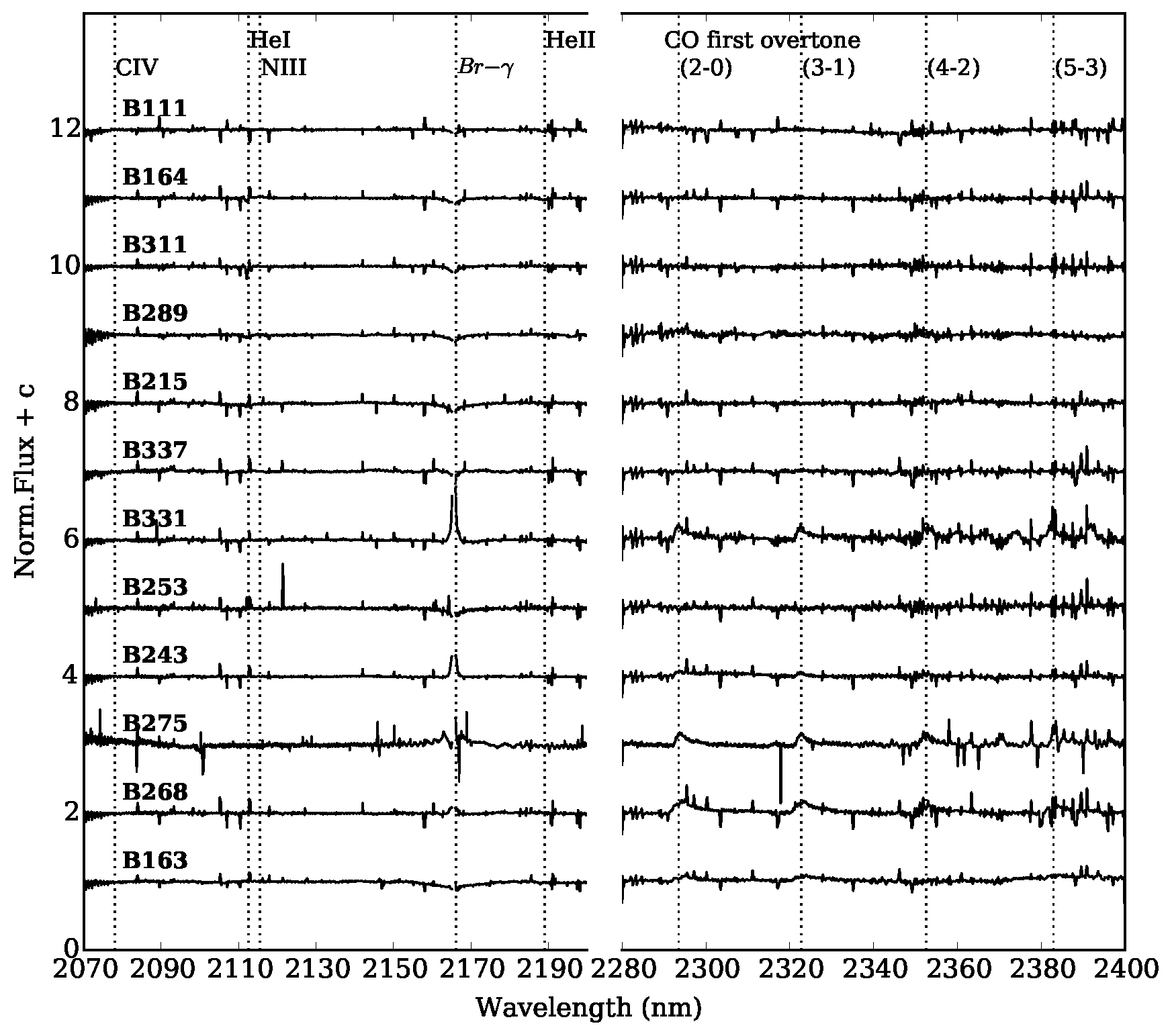}
      \caption{$K$-band spectra of our X-shooter targets. The location of the Br$\gamma$ line and CO bandhead emission are indicated. In B331, B243, B275 and B268 Br$\gamma$ is in emission. B163, B331, B243, B275 and B268 show CO bandhead emission. The narrow features in the spectrum are residuals of the sky subtraction procedure. The spectra have been smoothed to reduce the noise in the continuum and clipped to remove residuals of the sky subtraction in $\rm Br\gamma$.}
         \label{P1:fig:NIR}
   \end{figure*}

The flux calibration was obtained using spectrophotometric standards from the ESO database. We then scaled the NIR flux to match the absolutely calibrated VIS spectrum. The telluric correction was performed using the software tool \texttt{molecfit}  v1.2.0\footnote{\url{http://www.eso.org/sci/software/pipelines/skytools/molecfit}} \citep{2015A&A...576A..77S, 2015A&A...576A..78K}. Parts of the spectra are shown in Figs.~\ref{P1:fig:VisualSpec} to \ref{P1:fig:NIR} and the full spectrum is available in the on-line material of this paper. We used the \texttt{xsh\_scired\_slit\_nod} recipe to reduce the data, meaning that the sky subtraction is performed by subtracting the two different nodding positions. We note that due to the spatial variation of the nebular lines along the slit, residuals from the sky subtraction are present in some of the the reduced spectra (e.g., [\ion{N}{ii}] 6548,6584~\AA, [\ion{S}{ii}] 6716,6731~\AA\ and [\ion{S}{iii}] 9069,9532~\AA, and the \ion{H}{} lines). 

\section{Spectra and spectral classification}
\label{P1:Spectral classification}

Traditionally, the blue spectral region (400 -- 500~nm) is used to perform the spectral classification of OB-type stars. In the case of the M17 sources studied in this paper the blue region is severely affected by extinction, but is for most of our objects detected in the X-shooter spectrum (Fig.~\ref{P1:fig:VisualSpec}). Many targets show the full hydrogen Balmer, Paschen and Brackett (up to Br$\gamma$) series,  \ion{He}{i} and \ion{He}{ii} lines (the latter in the O-type stars), the Ca triplet lines, and diffuse interstellar bands. Figs.~\ref{P1:fig:CaT} and \ref{P1:fig:NIR} provide an overview of the Ca triplet region (820--880~nm) and the $K$-band (2100--2400~nm, including Br$\gamma$ and the CO bandheads) for all targets. 
With the exception of B163, B331 and B337 which are strongly affected by interstellar absorption in the blue ($A_V > 10$), we classified the stars based on the strength of the H, He and metal lines in the UVB spectrum. For the O stars we visually compared our spectra with the previously catalogued and published spectra by the Galactic O Star Spectroscopic Survey \citep[GOSSS;][]{2011ApJS..193...24S, 2014ApJS..211...10S, 2016ApJS..224....4M}. For the B stars we used the criteria and spectra published by \citet{2009ssc..book.....G}.

We looked for the presence of [\ion{O}{i}]\,6300. This forbidden line is associated with bi-polar outflows/jets, disk winds, and the disk surface in Herbig Ae/Be stars.  The [\ion{O}{i}] line is thought to originate from the region where the UV radiation from the star impinges on the disk surface \citep[][]{1985A&A...151..340F, 2008A&A...485..487V, 2011A&A...529A..34M}. We also paid special attention to the luminosity subclass Vz, which refers to objects with a substantially stronger \ion{He}{ii}\, 4686 line. These objects are hypothesised to be the youngest optically observable O type stars \citep[the 'z' stands for ZAMS;][]{2014A&A...564A..39S, 2016AJ....152...31A}. Given the young age of M17, one might expect such stars to be present.

Following this classification scheme we are able to determine the spectral types with an accuracy of one subtype, unless stated otherwise in Sect.~\ref{P1:ind_sources}. We constrained the luminosity class of the stars using the surface gravity derived in Sect.~\ref{P1:FASTWIND}. An overview of the spectral features described in this section is presented in Table~\ref{P1:tab:Summary_properties} and the results of the spectral classification are listed in Table~\ref{P1:tab:Stellar_parameters}. In most cases the spectral classification agrees well with the one provided in the literature by \citetalias{1997ApJ...489..698H} and \citet{2008ApJ...686..310H}.

\subsection{Individual sources}
\label{P1:ind_sources}

In this section we discuss the X-shooter spectra of the individual targets. In parentheses we list the updated spectral types corresponding to this work. The spectra are displayed in full in Appendix \ref{P1:appendix_spec}. 

\textbf{B111 (O4.5 V):} This clearly is an early O-type spectrum: the \ion{He}{ii} lines at 4200, 4541, and 4686~\AA\ are strong, consistent with an O4.5~V type star. The spectrum does not show \ion{He}{i}~4387, and \ion{He}{i}~4471 is weak. The hydrogen Balmer (including H$\alpha$) and Paschen series lines are strongly in absorption. Using its NIR spectrum, \citetalias{1997ApJ...489..698H} classified B111 as a kO3-O4 based on the \ion{He}{ii}, \ion{N}{ii} lines and the lack of C~{\sc iv}. According to our $K$-band spectrum B111 is a kO5-O6 type star, based on the presence of \ion{He}{ii} lines and weak, if any, C~{\sc iv} emission. The spectrum does not include any (nebular) emission lines.

\textbf{B163 (kA5):} Due to severe interstellar extinction it is not possible to identify any photospheric features in the UVB arm of the X-shooter spectrum. Longward of $\sim$8500~\AA\, the first Paschen line becomes detectable. The NIR spectrum also shows the Brackett series in absorption, no elements other than H are seen. Based on this spectrum we assign a mid-A spectral type. Two of the CO first overtone transitions are detected, indicating the presence of a circumstellar disk. The spectrum includes nebular emission lines in the hydrogen series, \ion{He}{i} lines (e.g., 10830~\AA), and [\ion{N}{iii}] 9069,9532~\AA. This points to an energetic ionising radiation field as expected given the location of B163 near the core of NGC~6618 (Fig.~\ref{P1:fig:M17rgb}). According to our criteria B163 is a mYSO. 

\textbf{B164 (O6 Vz):} The presence of \ion{He}{ii}~4541, 4686, and the fact that these lines are stronger than \ion{He}{i}~4387 and 4471 demonstrate that B164 is an O-type star. The ratio \ion{He}{ii}~4200~/~\ion{He}{i}~(+{\sc ii})~4026~\AA\ is near unity which indicates a spectral type O6~V. Based on the ratio EW(\ion{He}{ii}~4686) / EW(\ion{He}{ii}~4542)~$> 1.10$ we conclude that it belongs to the Vz class. From the $K$-band classification we conclude that B164 is a kO5-O6 type star. \citetalias{1997ApJ...489..698H}  classified B164 as kO7-O8 based on the \ion{N}{iii} and \ion{He}{i} lines. No emission lines are detected.

\textbf{B215 (IRS15; B0-B1 V):} No \ion{He}{ii} lines are detected, which indicates that B215 must be of later spectral type than B0. The hydrogen lines of the Balmer, Paschen and Brackett series are in absorption, and the \ion{He}{i} absorption lines are weak (e.g., 4922). The absence of \ion{Mg}{ii}~4481 suggests that the spectral type is earlier than B1; therefore, we conclude that this star is of spectral type B0-B1. The $K$-band classification is consistent with kO9-B1. \citet{2001A&A...377..273N}  classified this object as an extreme class I source, and \citet{2006ApJ...645L..61C} pointed out that it is a $\sim26$~$M_{\odot}$ star surrounded by an extensive remnant disk. The spectrum contains nebular emission lines in the hydrogen series, \ion{He}{i} lines and forbidden emission lines [\ion{N}{ii}] 6548,6584, [\ion{S}{ii}] 6716,6731 and [\ion{S}{iii}] 9069,9532. We also identify [\ion{O}{i}]\,6300 emission: a broad component ($\sim$100~km/s) on top of a nebular component. By inspecting the 2D frames we conclude that the emission seen in H$\alpha$ has a nebular origin. From the X-shooter spectrum alone we cannot conclude that this is a mYSO.

\textbf{B243 (B8 V):} The hydrogen lines of the Balmer series are in absorption. The strongest Balmer lines show a central emission component; H$\alpha$ is strongly in emission. The Paschen series absorption lines are filled in with double-peaked emission lines. The Brackett series lines mainly exhibit a double-peaked emission component. Furthermore, the `auroral' \ion{O}{i} 7774 and 8446 lines show prominent double-peaked emission. The \ion{Ca}{ii} triplet is not present. The spectrum includes \ion{He}{i} absorption lines (e.g. 4471, 5876, 10830); the ratio \ion{He}{i}~4026~/~\ion{He}{i}~4009 is around two. The \ion{He}{i}~4471 and \ion{Mg}{ii}~4481 line ratio is close to unity which points towards spectral type B8. The $JHK$ CCD indicates NIR excess, weak CO-bandhead emission is detected, and it presents [\ion{O}{i}]\,6300 emission. Therefore we conclude that B243 is a mYSO.

\textbf{B253 (B3-B5):} The spectrum displays strong and broad hydrogen absorption lines, with a central (nebular) emission component. \ion{He}{i} absorption lines are present (e.g., 4026, 4471, 4922, 10830~\AA), also with a central (nebular) emission component; the \ion{He}{i} lines are most prominent in the blue part of the spectrum. No \ion{He}{ii} lines are detected, nor the \ion{C}{ii}~4267~\AA\ line; therefore, we conclude that B253 is a B3-B5 type star. The NIR CMD and CCD show no evidence for a NIR excess. Its $K$-band spectrum is kB5 or later. Many forbidden emission lines are present (e.g., [\ion{O}{iii}], [\ion{N}{ii}], [\ion{S}{ii}]) that, together with the \ion{He}{i} emission, indicate a high degree of ionisation of the surrounding \ion{H}{ii} region.

\textbf{B268 (B9-A0):} The hydrogen Balmer and Paschen series are prominently in absorption and include a central nebular emission component. The H$\alpha$ and H$\beta$ profiles are filled in by a circumstellar emission component. The \ion{He}{i} lines are very weakly present, the nebular emission component dominates the line profiles. The fact that \ion{He}{i}~4471 and \ion{Mg}{ii}~4481 have almost the same strength indicates that the spectral type is late-B to early-A. The Ca triplet lines show double-peaked emission and a red-shifted absorption component that reaches below the continuum and could indicate the presence of an accretion flow \citep{1994ApJ...426..669H}. We confirm the findings of \citetalias{1997ApJ...489..698H} who classified B268 as a mYSO, based on the presence of CO bandhead emission and the Pa$\delta$ line.

\textbf{B275 (B7 III):}  \citet{ 2011A&A...536L...1O} performed a detailed spectral classification, resulting in spectral type B7~III. They concluded that B275 is a pre-main-sequence star contracting towards the main sequence. The  \ion{He}{i}~4009~\AA\ and \ion{C}{ii}~4267~\AA\ lines are very weak and when considering the Si\,{\sc ii}~4128~\AA\ to Mg\,{\sc ii}~4481~\AA\ ratio the spectral type becomes B6-B7. The \ion{O}{i}~8446~\AA\ and Ca triplet lines show pronounced double-peaked emission, as do several of the hydrogen lines on top of a photospheric absorption profile. Both first- and second-overtone CO emission is detected. The [\ion{O}{i}]\,6300 emission line is present. Together with double-peaked emission features and NIR excess, this points to the presence of a rotating circumstellar disk. B275 is a massive YSO.

\textbf{B289 (O9.7 V):} The spectrum includes \ion{He}{ii} 4686 and \ion{He}{ii} 5411 absorption lines, but the presence of \ion{He}{ii}~4200 is hard to confirm given the low signal-to-noise ratio. The \ion{He}{i}~(+{\sc ii})~4026 line is present, \ion{He}{i}~4144 and 4387 are weak and \ion{He}{i}~4471 and 5876 are strong. In addition, it is possible to identify the C\,{\sc iii}~4647/50/51 line complex with a similar strength as the \ion{He}{ii}~4686 line, so that the spectral type is O9.7~V. From the $K$-band we obtain a kO9-B1 spectral type. According to \citetalias{1997ApJ...489..698H}, B289 might have a NIR excess and be a late O star. The $JHK$ CCD does not display evidence for a NIR excess and there are no emission lines, therefore we cannot conclude that it is a mYSO. The  1.5 and 2.0~$\mu$m photometry by \citet{2001A&A...377..273N} using MANIAC mounted at the ESO La Silla 2.2~m telescope shows IR excess. It is one of the sources surrounded by an IR-bright dusty disk as reported by \citet{2005IAUS..227..145C}. The spectrum shows some weak nebular emission lines.
 
\textbf{B311 (O8.5 Vz):} The \ion{He}{i}~(+{\sc ii})~4026 and  \ion{He}{i}~4471 lines are almost equally strong and sharp.  \ion{He}{i}~$\lambda \lambda$~4144, 4387, and \ion{He}{ii}~$\lambda \lambda$~4541, 4686 also show up in the spectrum. The \ion{C}{iii}~4068/69/70 complex is in absorption together with \ion{Si}{iv}~4089 and 4116. We classify this star as O8.5~Vz because the ratio EW(\ion{He}{ii}~4686)\,/EW(\ion{He}{i}~4471)~$>$~1.10. From the NIR spectrum, \citetalias{1997ApJ...489..698H} classified it as later than O9-B2 because it has \ion{He}{i} in absorption and lacks \ion{N}{iii} and \ion{He}{ii}. \citet{2001A&A...377..273N} and \citet{2005IAUS..227..145C} detected an IR-bright dusty disk in the $N$ and $Q$-bands. However, \citet{2008ApJ...689..242P} resolved a bow shock associated with it, which is responsible for the NIR emission. B311 is a main-sequence star according to our criteria. The result of the $K$-band classification agrees well with the visual spectral type (kO9-B1). Nebular emission is weakly present. 

\textbf{B331 (late-B):} The UVB part of the spectrum is not detected due to the severe interstellar extinction towards this source. The red and near-infrared spectrum is dominated by strong emission lines. In the case of the hydrogen series lines these are superposed on a broad and shallow photospheric profile. According to \citet{2008ApJ...686..310H}, B331 is a B2~V type star based on its visual spectrum. No helium lines are detected in the NIR part of the spectrum, indicating that B331 is a late-B or early-A-type star. It exhibits Br$\gamma$, \ion{O}{i}, CO bandhead emission, and several double-peaked emission lines indicating the presence of a rotating circumstellar disk. The SED includes a NIR excess, making it a bonafide mYSO. The spectrum contains some weak nebular emission lines. 
 
\textbf{B337 (late-B):} This object is deeply embedded so that it is not possible to detect the blue part of the spectrum. It does show the Paschen series in absorption with a narrow central nebular emission component. The Ca triplet lines show pronounced double-peaked emission and a blue absorption component remnant of the sky subtraction. The SED presented by \citetalias{1997ApJ...489..698H} is consistent with a B5~V type star when correcting for the distance. No \ion{He}{i} lines can be identified in the X-shooter spectrum pointing towards a late-B or early-A-type star. The NIR CCD indicates a NIR excess and the \ion{Ca}{ii} triplet lines are in emission and double-peaked; we do not detect CO bandhead emission. B337 is a mYSO.

\begin{table}
\centering
\caption{\normalsize{Summary of the properties of the sources observed with X-shooter.}}
\begin{minipage}{\hsize}
\centering
\renewcommand{\arraystretch}{1.4}
\begin{tabular}{lcccc}
\hline
\hline
Source & Emission & CO  & NIR & Classification \\
& lines & bandhead & excess & \\
 \hline B111 	&	- 	&	-	&	-  & O star   \\
B163    &   -   &   +   &   +   & mYSO   \\
B164 	&	- 	&	-  	&	-   & O star   \\
B215    &   -   &	-	&	*\footnote{\label{P1:foot:NIRex}NIR excess only longwards of 3~$\mu$m, not in the X-shooter wavelength range.} &   B star? \\
B243 	&	+	&	+	&	+   & mYSO   \\
B253 	&	-	&	-	&	-   & B star   \\
B268 	&	+	&	+	&	+   & mYSO   \\
B275 	&	+	&	+	&	+   & mYSO   \\
B289 	&	- 	&	-	&	*\footref{P1:foot:NIRex}       & O star?\\
B311 	&	- 	&	-	&	-  & O star   \\
B331 	&	+ 	&	+	&	*\footref{P1:foot:NIRex}  & mYSO    \\
B337 	&	+ 	&	-	&	+  & mYSO    \\
\hline
\vspace{-15pt}																	
\end{tabular}
\renewcommand{\footnoterule}{}
\end{minipage}
\label{P1:tab:Summary_properties}																	
\end{table}

\section{Spectral energy distribution}
\label{P1:SEDs}

\begin{figure*}
\setlength{\tabcolsep}{2pt}
     \centering
        \subfigure[M17-B111]{%
          \label{P1:fig:SED111}
          \includegraphics[width=5.5cm]{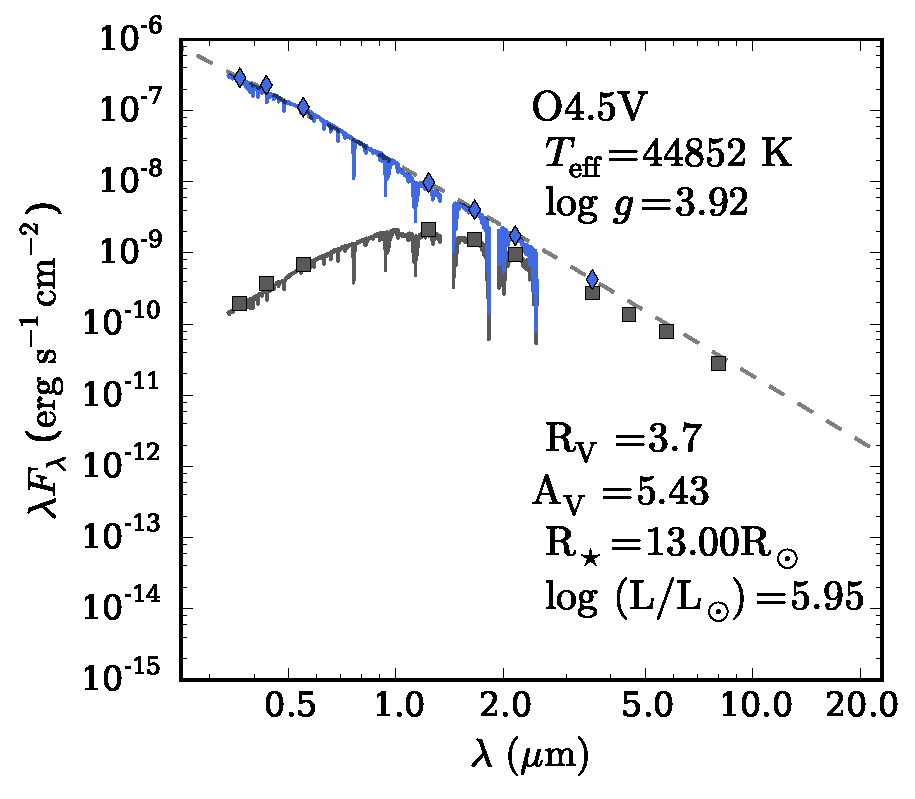}
        }%
        \subfigure[M17-B163]{%
           \label{P1:fig:SED163}
           \includegraphics[width=5.5cm]{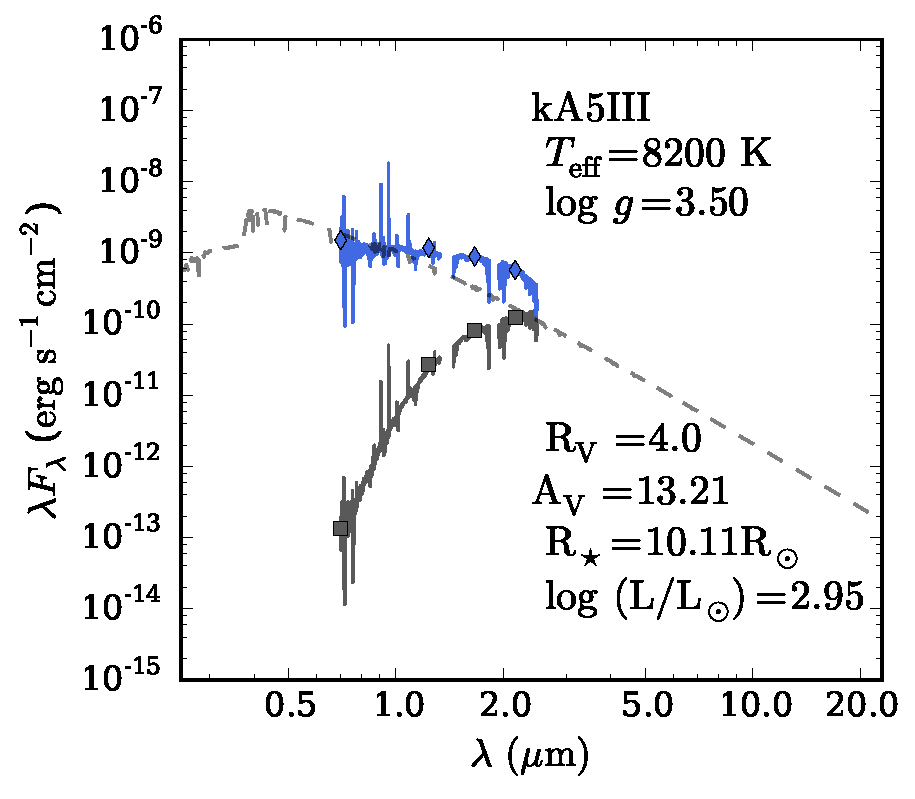}
        }%
          \subfigure[M17-B164]{%
            \label{P1:fig:SED164}
            \includegraphics[width=5.5cm]{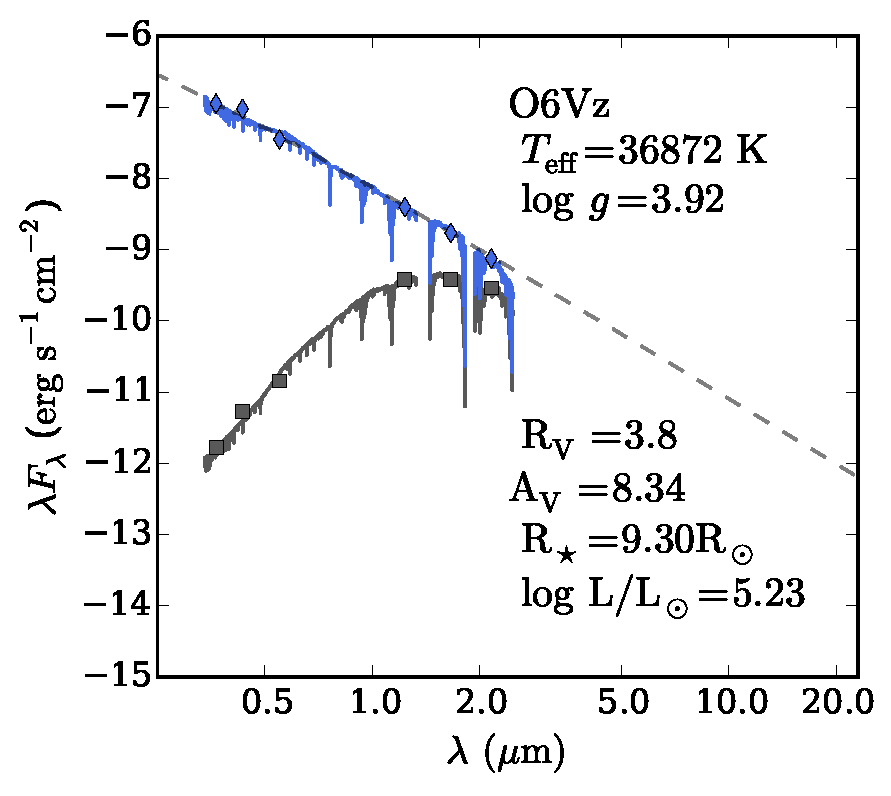}
        }\\%
        \subfigure[M17-B215 (IRS15)]{%
           \label{P1:fig:SED215}
           \includegraphics[width=5.5cm]{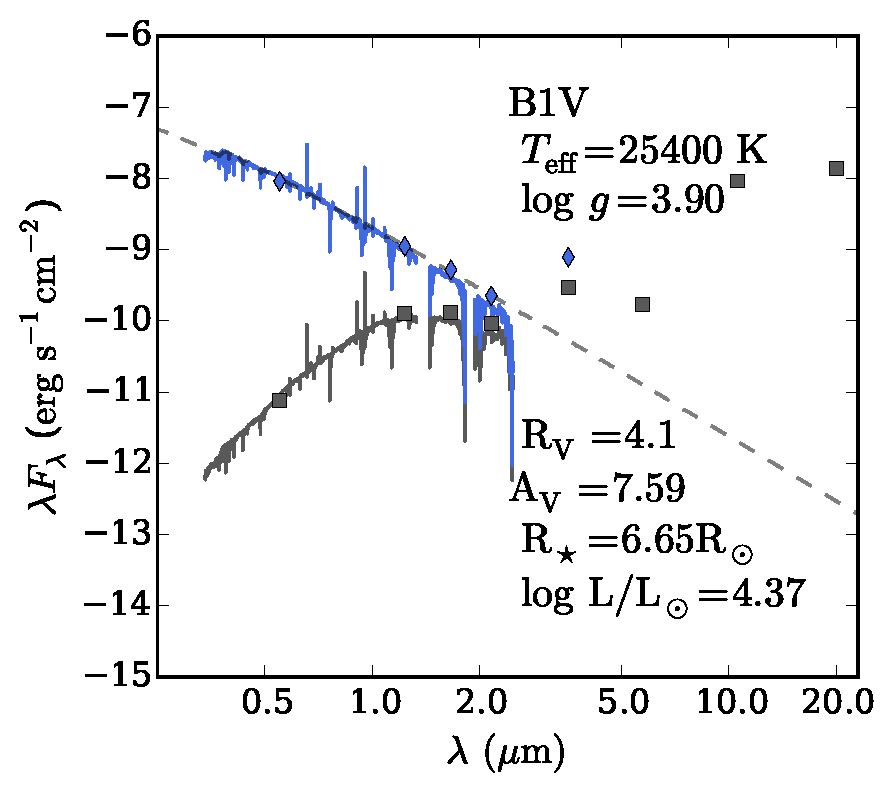}
        }%
        \subfigure[M17-B243]{%
           \label{P1:fig:SED243}
           \includegraphics[width=5.5cm]{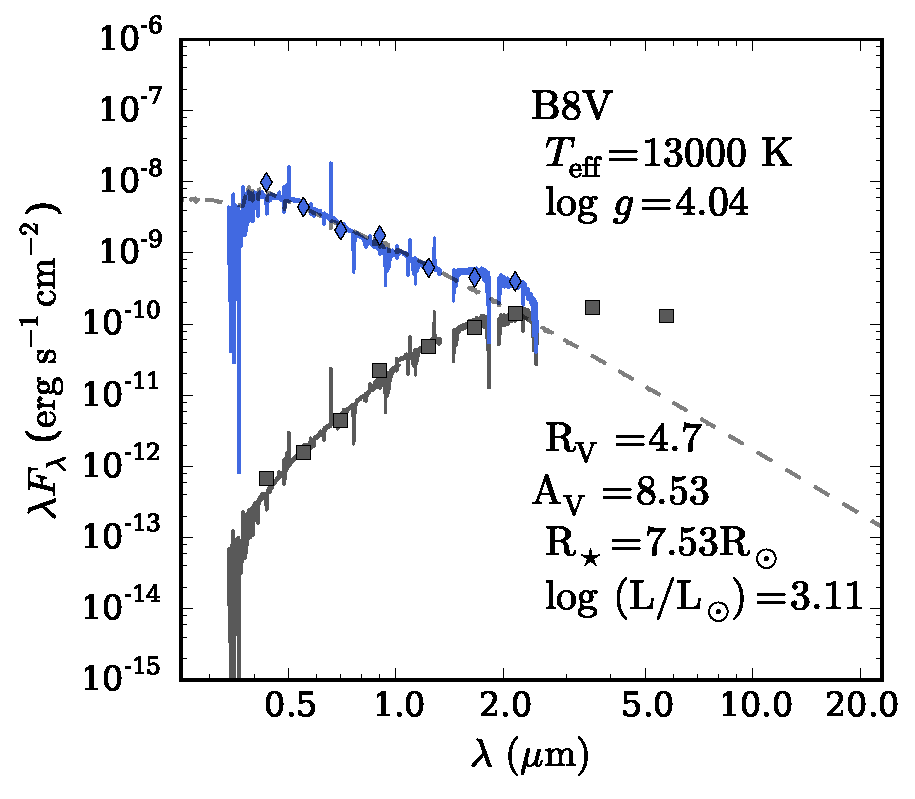}
        }%
          \subfigure[M17-B253]{%
            \label{P1:fig:SED253}
            \includegraphics[width=5.5cm]{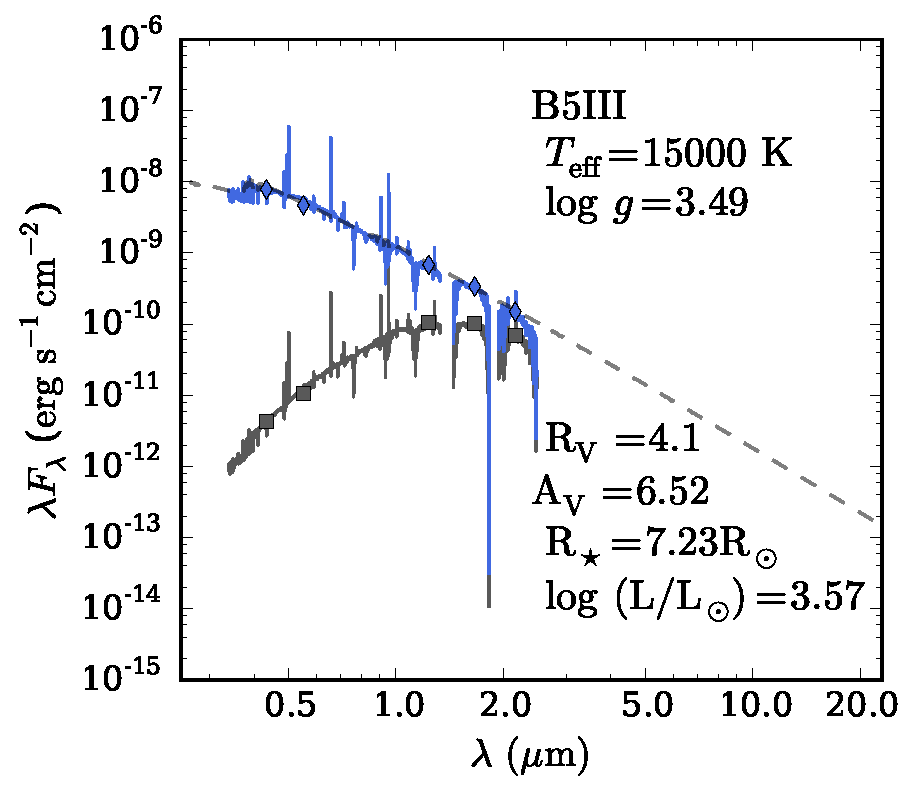}
        }\\%
      	   \subfigure[M17-B268]{%
            \label{P1:fig:SED268}
            \includegraphics[width=5.5cm]{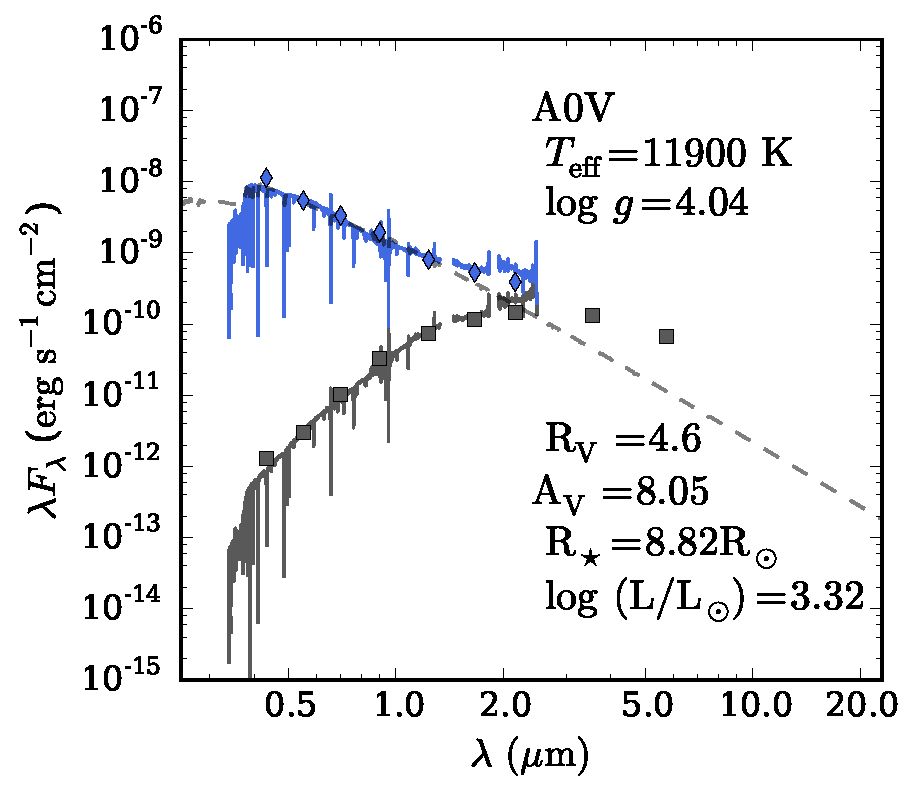}
        }%
        \subfigure[M17-B275]{%
           \label{P1:fig:SED275}
           \includegraphics[width=5.5cm]{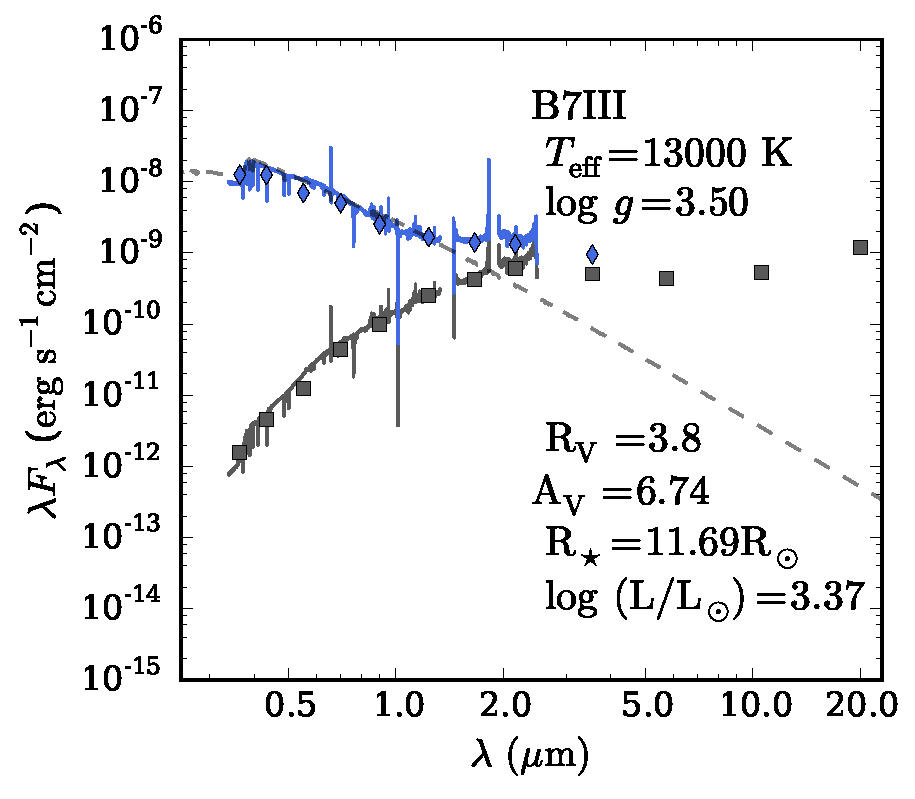}
        }%
          \subfigure[M17-B289]{%
            \label{P1:fig:SED289}
            \includegraphics[width=5.5cm]{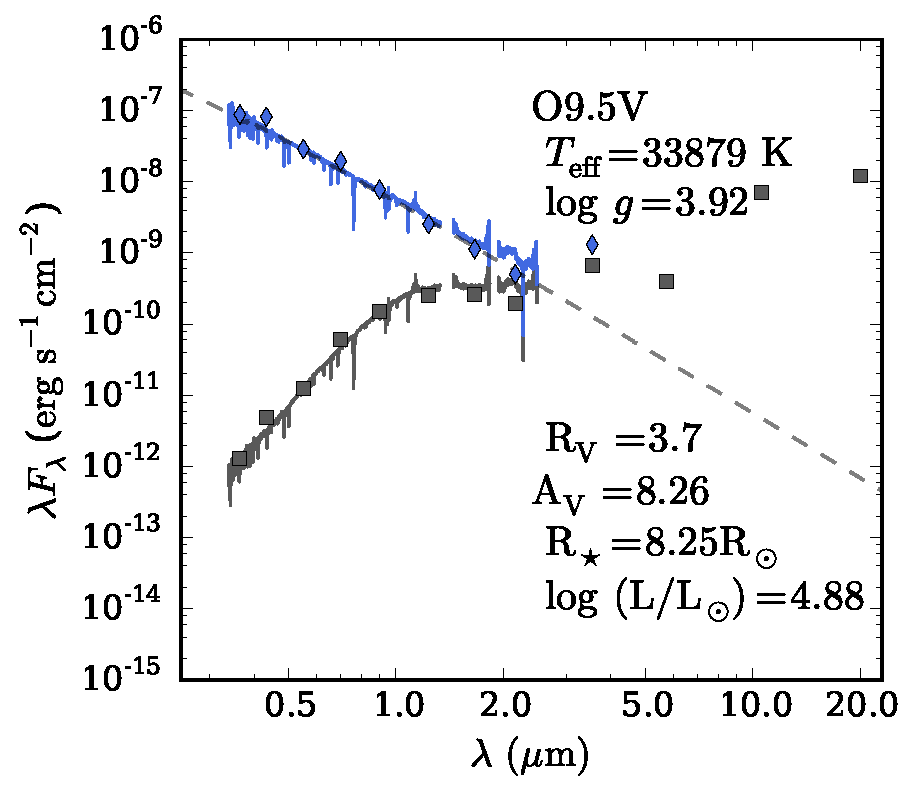}
        }\\%
           \subfigure[M17-B311]{%
            \label{P1:fig:SED311}
            \includegraphics[width=5.5cm]{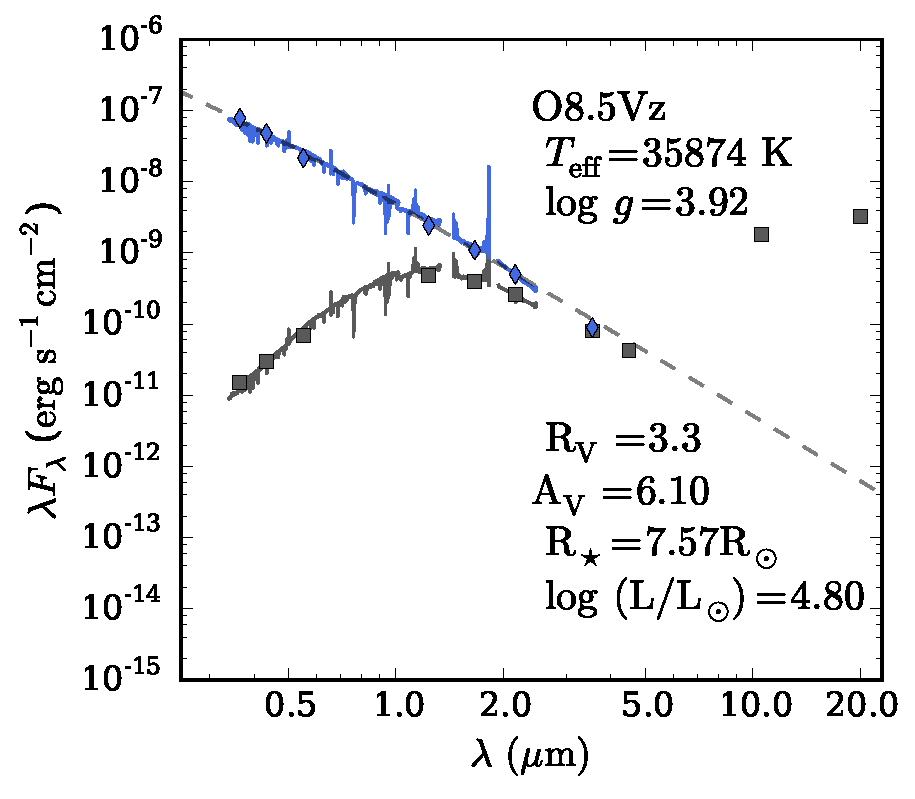}
        }%
        \subfigure[M17-B331]{%
           \label{P1:fig:SED331}
           \includegraphics[width=5.5cm]{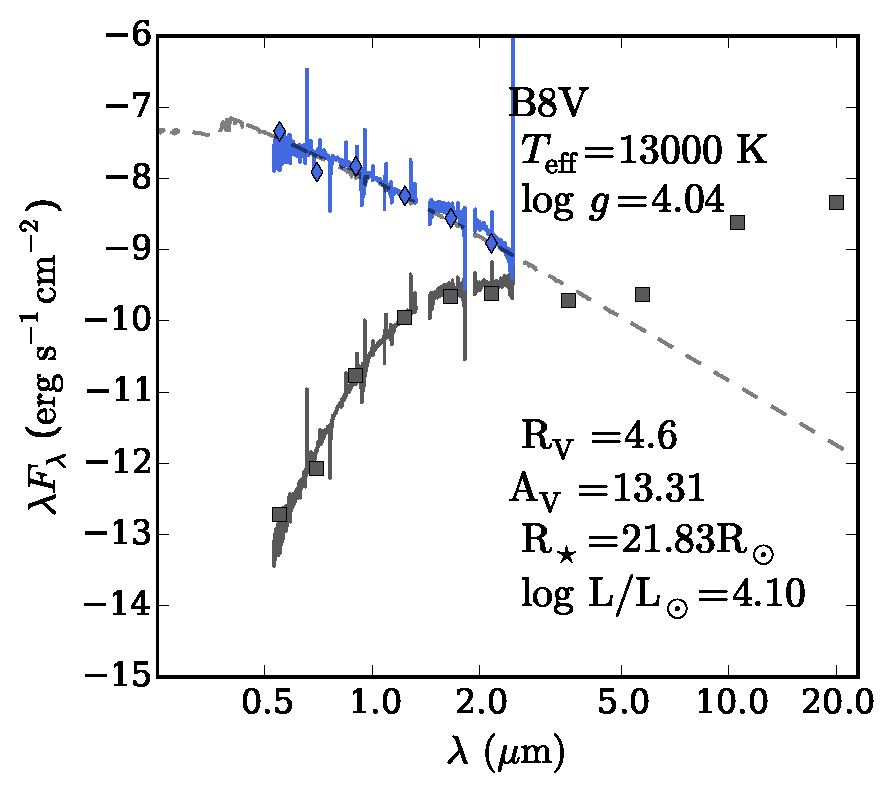}
        }%
        \subfigure[M17-B337]{%
           \label{P1:fig:SED337}
           \includegraphics[width=5.5cm]{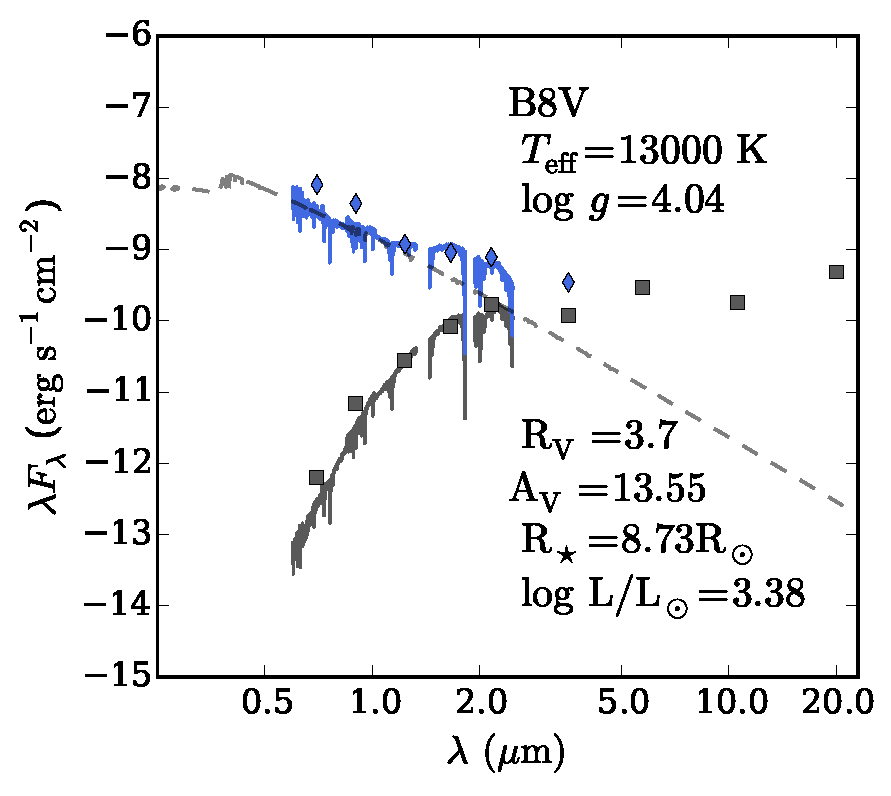}
        }%

    \caption[]{Spectral energy distributions of our X-shooter targets. The dashed line shows the Castelli \& Kurucz model, the blue line the dereddened X-shooter spectrum, and the solid black line the flux calibrated spectrum. The squares and diamonds give the observed and dereddened magnitudes, respectively. In the top part of the plots we indicate the spectral type of the star and the parameters of the corresponding Castelli \& Kurucz model. In the bottom part of the plots we list the extinction parameters, and the radius and luminosity of the star resulting from our SED fit. The spectra of B163, B331, and B337 were clipped in order to show only the part of the spectrum used for the analysis.}
\label{P1:fig:SEDs}
\end{figure*}

We construct the spectral energy distribution (SED) of the star by dereddening the X-shooter spectrum, as well as the available photometric data (extending into the mid-infrared wavelength range). We fit the slope of the SED in the photospheric domain (400 -- 820~nm) to Castelli \& Kurucz models \citep{1993yCat.6039....0K, 2004astro.ph..5087C}. As an initial guess, we took the associated $T_{\rm eff}$ and $\log\,g$ from the Castelli \& Kurucz\footnote{Table 2 in \url{http://www.stsci.edu/hst/observatory/crds/castelli_kurucz_atlas.html}} model corresponding to the spectral type reported in Sect.~\ref{P1:Spectral classification}. We then cross-check our choice of $T_{\rm eff}$ and $\log\,g$ with the results obtained in Sect.~\ref{P1:FASTWIND} and perform the necessary iterations. The final values of $T_{\rm eff}$ and $\log\,g$, corresponding to the best fit Castelli \& Kurucz models used, are displayed in Figure~\ref{P1:fig:SEDs}. We thereby constrain the visual extinction $A_V$ and the total to selective extinction $R_{V} = A_V/$E$(B-V)$ towards the objects, adopting the parametrisation of \citet{1989ApJ...345..245C}. From comparing these SEDs to observed spectra a near-infrared excess should become apparent (Figure \ref{P1:fig:SEDs}).

The flux calibrated X-shooter spectra and the photometric data points are shown by the black lines and squares, respectively. The photometric data in the $UBVRI$ bands were taken from \citet{1980A&A....91..186C} who used the 123~cm telescope on Calar Alto, and the 50~cm telescope on the Gamsberg in South West Africa. These measurements have an accuracy of $\pm$0.5~mag. We highlight that the apertures used for this study are quite large, therefore the measurements might be contaminated by close neighbours. For B215 and B331 we obtained the $V$-band magnitude from \citet{2008ApJ...686..310H}, and for B111 from the AAVSO Photometric all sky survey (APASS)\footnote{\url{https://www.aavso.org/apass}}. The $JHK$ magnitudes were  taken from the 2MASS survey \citep{2006AJ....131.1163S} which provides photometry for our objects with an accuracy better than 0.3~mag, with the exception of B215 whose $JHK$ photometry we obtained from \citet{1998A&A...329..161C}. The mid-IR magnitudes were measured with the Infrared Array Camera (IRAC) on the \emph{Spitzer} Space Telescope \citep{2004ApJS..154...10F} as part of the \emph{Spitzer} Legacy Science Program GLIMPSE \citep{2003PASP..115..953B}; the photometric accuracy achieved for our targets is better than 0.2~mag. The $N$ and $Q$-band magnitudes (10.5 and 20 $\mu$m) were obtained (when available) from \citet{2001A&A...377..273N} who observed M17 with the ground-based infrared camera MANIAC. The aperture sizes were 2" for B337, 3" for B331 and B275, 5" for B311, 7" for B215, and 10" for B289; the accuracy in their measurements varies from 10 to 30\%. The dereddened spectra and photometry are shown in blue. The dashed grey line represents the Castelli \& Kurucz model corresponding to the spectral type of the star. The temperature and $\log{g}$ corresponding to the best Castelli \& Kurucz model is indicated in the bottom-right corner; note that this is not identical to the temperature and $\log{g}$ of the star, but it corresponds to the adopted Castelli \& Kurucz model. The resulting extinction and stellar parameters are labeled as well.

\subsection{Extinction parameters}
\label{P1:Extinction}

In order to obtain independent values of the total extinction $A_V$ and the total to selective extinction $R_V$, we implemented a $\chi^2$ fitting algorithm. We first dereddened the X-shooter spectrum varying $R_V$ from 2 to 5.5 in steps of 0.1 and then fitted the slope of the Castelli \& Kurucz model to that of the dereddened X-shooter spectrum in the photospheric domain to determine the total V-band extinction, $A_V$.
This allowed us to constrain $R_V$ for all sources, except for B163. For this star we lack spectral coverage at $\lambda < 850$~nm. As the infrared excess starts at 1000~nm, a too limited spectral range is available for constraining $R_V$.

The obtained values for $R_V$ range from 3.3 to 4.7 and $A_V$ varies from $\sim$6 to $\sim$14~mag (Table~\ref{P1:tab:Stellar_parameters}). \citetalias{1997ApJ...489..698H} observe a similar range, while \citet{2008ApJ...686..310H} found $R_V = 3.9 \pm 0.2$ for their sample. The latter authors argue that the extinction to M17 is best described by a contribution of foreground extinction ($A_V = 2$~mag with $R_V = 3.1$, the average Galactic value) plus an additional contribution produced by local ISM dust. We derive $R_V$ and $A_V$ for each individual sight-line. The $A_V$ values that we obtained agree within the errors with those calculated by \citet{2009ApJ...696.1278P} from the total hydrogen column density $N$(H).

The X-shooter spectra include strong diffuse interstellar bands (DIBs). \citetalias{1997ApJ...489..698H} reported that their strength did not vary with $A_V$ (or E($B-V$)) as seen in other Galactic sightlines. Massive star forming regions are known to exhibit anomalous extinction properties \citep{2013ApJ...773...41D, 2013ApJ...773...42O, 2013A&A...558A.102E}. We will report the DIB behaviour towards sightlines in M17 in a separate paper.

\subsection{Stellar radius}
\label{P1:radius}

To estimate the radius of the stars we scaled the observed flux to the flux produced by the stellar surface given by the Castelli \& Kurucz model, using the distance to M17 ($d=1.98$~kpc). The difference between the measured flux $\rm F_{\lambda}$ and the flux from the model $\rm F_{Kur}$ can be corrected for by multiplying by a factor $(R_{\star}/{d})^2$, where $R_{\star}$ is the stellar radius, and $d$ is the distance to the Sun. We calculated the luminosity $ \log{L/L_{\sun}}$ using the $V$-band magnitude from \citet{1980A&A....91..186C}, the distance to M17, the $A_V$ values from our fit (Sect.~\ref{P1:Extinction}), and the bolometric correction corresponding to the spectral type from \citet{2000asqu.book.....C}.

To assess the uncertainties in the parameters, we calculate the probability $P$ that its $\chi^2$ value differs from the best-fit $\chi^2$ due to random fluctuations: $P=1-\Gamma(\chi^2 /2,\nu /2)$, where $\Gamma$ is the incomplete gamma function and $\nu$ the number of degrees of freedom. We normalise the $\chi^2$ such that the reduced $\chi^2$ corresponding to the best fitting model has a value of unity and we select all models with $P \geq 0.32$ as acceptable fits representing the 68\% confidence interval. The finite exploration of the parameter space may result in an underestimation of the confidence interval near the borders of $P(\chi^2,\nu)$ = 0.32. To avoid an underestimation of the errors we select as boundaries of the 68\% confidence interval the first combination of parameters that do not satisfy $P(\chi^2,\nu) \geq 0.32$ \citep[see ][]{2014A&A...572A..36T, {2017arXiv170104758R}}. Like in the case of B275 \citep{2011A&A...536L...1O} the obtained stellar radius of the mYSOs does not correspond to the value expected for a main-sequence star. This corroborates our earlier finding that these mYSOs indeed are massive pre-main-sequence stars that are still contracting towards the main sequence. In Sect.~\ref{P1:FASTWIND} we will show that this is consistent with the $\log{g}$ values independently measured from the broadening of the diagnostic lines.

\section{Modeling the photospheric spectrum}
\label{P1:FASTWIND}

To further constrain the temperature, luminosity, projected rotational velocity, and surface gravity of the stars we used an automatic fitting algorithm developed by \citet{2005A&A...441..711M} and \citet{2017arXiv170104758R}, which compares the H\,{\sc i}, \ion{He}{i}, \ion{He}{ii}, and N absorption lines with model profiles produced by the non-LTE stellar atmosphere model \texttt{FASTWIND} \citep{2005A&A...435..669P, 2012A&A...537A..79R}. For our PMS stars we did not fit the He and N abundances but fixed them to be consistent with the solar values. This is in agreement with what we would expect for such a young stellar population. This method applies the genetic fitting algorithm \texttt{PIKAIA} \citep{1995ApJS..101..309C} which allows us to explore the parameter space in an extensive way \citep{2005A&A...441..711M, 2014A&A...572A..36T}. \texttt{FASTWIND} calculates non-LTE line-blanketed stellar atmospheres and accounts for a spherically symmetric stellar wind. It can be used to examine the dependence of H, \ion{He}{i}, and \ion{He}{ii} photospheric lines on $T_{\rm eff}$ and $\log{g}$. Its application to B stars has been successfully tested by \citet{2010A&A...515A..74L}.

Inputs to the fit are the absolute $V$-band magnitude ($M_{V}$) and the radial velocity (RV) of the star. We calculated $M_{V}$ using the apparent magnitude reported by \citet{1980A&A....91..186C} and the extinction coefficient $A_V$ found in Sect.~\ref{P1:SEDs} adopting a distance of 1.98~kpc. RV was measured following the procedure described by \citet{2013A&A...550A.107S} where we simultaneously adjust the spectral lines for a given object, taking into account variations in the signal-to-noise ratio. The RV values vary from -11 to 20~$\rm km$\,$\rm s^{-1}$ and are listed in Tab.~\ref{P1:tab:Stellar_parameters}; for B243, B268, and B289 we list the RV values obtained from the 2013, 2013, and 2012 observations, respectively. We calculated the RV dispersion using the standard deviation of a gaussian distribution ($\sigma_{1D}$) and we find the strikingly low value of $\sim$5~km\,$\rm s^{-1}$. This is in contrast with the expectation that the massive star population in M17 includes many (close) binaries \citep[e.g.][]{2007A&amp;A...474...77K, 2012Sci...337..444S, 2015A&A...580A..93D}. This points to a lack of short period binaries or a low binary fraction. A quantitative investigation of the low $\sigma_{1D}$ is presented in \citet{2017arXiv170202153S}.

Several atmospheric parameters are obtained from the fitting procedure. The effective temperature ($T_{\rm eff}$) can be constrained using the relative strength of the H\,{\sc i}, \ion{He}{i}, and \ion{He}{ii} lines. The surface gravity ($\log{g}$) is obtained from fitting the wings of the Stark-broadened H\,{\sc i} lines. This parameter allows us to constrain the luminosity class \citep[see][]{2005A&A...441..711M}. The projected rotational velocity ($v\sin{i}$) is a natural outcome of the fitting procedure as the models are convolved with a rotational profile to reproduce the observed spectrum. The mass-loss rate is mainly determined by fitting the H$\alpha$ line. Only for B111 ($\log{\dot{M}} = -6.00 \pm 0.1$) and B164 ($\log{\dot{M}} = -6.35 \pm 3.65$), the two hottest stars in our sample, we obtain a reliable measurement. The parameter describing the rate of acceleration of the outflow ($\beta$) cannot be constrained by our data and was therefore fixed to the theoretical value predicted by \citet{2012A&A...537A..37M}, i.e. $\beta$ = 0.8 for main-sequence stars. Given the low mass-loss rates, the value adopted for $\beta$ does not affect the model results. Another parameter in the fitting procedure is the microturbulent velocity: we allowed this parameter to vary from 5 to 50\,$\rm km$\,$\rm s^{-1}$, but due to the lack of Si lines in our spectra this parameter remains poorly constrained. This does not affect the determination of the other stellar parameters.

The bolometric luminosity ($\log{L/L_{\odot}}$) is obtained by applying the bolometric correction to the absolute magnitude used as input. The luminosity and effective temperature are used to calculate the radius of the star $R_{\star}$. Using $\log{g}$ we can calculate the spectroscopic mass of the star $M_{\rm spec}=gR^2/G$, where $G$ is the gravitational constant. Given the uncertainty in $\log{g}$ it is very difficult to constrain $M_{\rm spec}$ as is evident from Tab.~\ref{P1:tab:Stellar_parameters}.

To calculate the errors in the parameters we follow the procedure described by \citet{2014A&A...572A..36T} and \citet{2017arXiv170104758R}. The best-fitting model is selected based on the $\chi^2$ value in the same way as described in Sect.~\ref{P1:radius}. The normalisation of the $\chi^2$ in this approach is only valid if the best fit is a good representation of the data, which we checked visually for each case (see Appendix~\ref{P1:appendix_fastwind}).
We select all models with $P(\chi^2, \nu) \geq 0.05$ as acceptable fits, representing the 95\% confidence interval to the fitted parameters. 
The results from the fitting procedure are listed in Tab.~\ref{P1:tab:Stellar_parameters}; the temperatures obtained with \texttt{FASTWIND} are consistent with the spectral types derived in Sect.~\ref{P1:Spectral classification}. The best fitting model, and other acceptable fits (5\% significance level or higher models) are shown in Appendix~\ref{P1:appendix_fastwind}. The lines used for each fit are displayed in figures~\ref{P1:fig:FASTWIND_fit_B111_B164} to~\ref{P1:fig:FASTWIND_fit_B311}; we have given the same weight to all the lines in the fitting procedure.

It is important to point out that the confidence intervals cited in this paper represent the validity of the models as well as the errors on the fit. They do not include the contribution from systematic errors due to the model assumptions, continuum placement biases, etc. For a detailed analysis of the systematic errors and their impact on the parameter determination the reader is referred to \citet{2017arXiv170104758R}. As can be seen in Tab.~\ref{P1:tab:Stellar_parameters} the error bars differ significantly from star to star. Large error bars are obtained in two cases: (i) when the parameter space is poorly explored near the border of the confidence interval and, therefore, the first model that does not satisfy $P(\chi^2, \nu) = 0.05$ is considerably outside the confidence interval; (ii) when the parameter is poorly constrained due to e.g., a low signal-to-noise ratio.

For B275 a good fit of the spectrum could not be obtained while leaving all parameters free. To mitigate this, we first constrained $v\sin\,{i}$ using a fit to only the helium lines. The obtained range of valid values was then used in a fit including the full set of diagnostic lines. H$\beta$ was not included in this fit because of the presence of a very strong 4880 \AA\ DIB blending the red wing of the line. This approach results in an acceptable fit to the spectrum, consistent with the results presented in \citet{2011A&A...536L...1O}; we note that the red wings of \ion{He}{i}~6678 and 5875 are not well represented. The morphology of these lines and other \ion{He}{i} lines could be an indication that this star is in a binary system; follow-up observations of this source are required to confirm or reject this hypothesis.

\begin{table*}
 \centering 
 \caption{\normalsize{Stellar properties derived from best-fit \texttt{FASTWIND} parameters and the SED fitting. The values for the stellar radius ($\rm R_{\star}$) cited here are the ones obtained via SED fitting. The sources classified as PMS stars in this work are in bold face, and the two sources for which we cannot confirm the PMS nature are underlined.}}
\begin{minipage}{\textwidth}
 \centering 
\renewcommand{\arraystretch}{1.4}
\setlength{\tabcolsep}{5pt}
\begin{tabular}{cccllllllllc}
 \hline 
 \hline 
Name &  Sp. Type & Sp. Type & \multicolumn{1}{c}{$T_{\rm eff}$} & $\log{g}$ & $v\sin{i}$ & $R_V$ & $A_V$ & $\log{L/L_{\odot}}$ & $R_{\star}$  & $M_{\rm spec}$ & RV \\
& Hanson & This work & \multicolumn{1}{c}{K} & cm\,$\rm s^{-2}$ & km\,$\rm s^{-1}$ & (SED) & mag & & $R_{\odot}$  & $M_{\odot}$ & km\,$\rm s^{-1}$ \\
\hline 

B111	 & 	O5 V	 & 	O4.5 V	 & 	$42850_{-1600}^{+3050}$ & $3.82_{-0.27}^{+0.13}$ & $170_{-28}^{+42}$ &$3.7_{-0.3}^{+0.4}$ &$5.4_{-0.3}^{+0.4}$ &$5.71_{-0.04}^{+0.08}$ &$13.0_{-0.6}^{+0.4}$ &$42_{-13}^{+12}$ & 3.4$\pm$0.8\\ 
\textbf{B163}	 & 	?	 & 	kA5	 & 	$8200$ & $-$ & $-$ & $4.0_{\downarrow}^{\uparrow}$ &$13.2_{\downarrow}^{\uparrow}$ &$2.95_{\downarrow}^{\uparrow}$\footnote{\label{P1:foot:SED}Values from SED fitting} &$10.1_{\downarrow}^{\uparrow}$ & $-$  & $-$\\ 
B164	 & 	O7-O8	 & 	O6 Vz	 & 	$39100_{-2350}^{+1150}$ & $4.01_{-0.27}^{+0.16}$ & $108_{-22}^{+30}$ &$3.8_{-0.3}^{+0.3}$ &$8.3_{-0.5}^{+0.4}$ &$5.09_{-0.07}^{+0.03}$ &$9.3_{-0.2}^{+0.4}$ &$22_{-8}^{+10}$  & -1.8$\pm$2.3\\ 
\underline{B215}	 & 	B0.5 V	 & 	B0-B1 V	 & 	$23500_{-3450}^{+6100}$ & $3.67_{\downarrow}^{+0.5}$ & $208_{-38}^{+40}$ &$4.1_{-0.1}^{+0.2}$ &$7.6_{-0.1}^{+0.3}$ &$3.98_{-0.15}^{+0.23}$ &$6.7_{-1.04}^{+0.94}$ &$6_{-2}^{+9}$ & 12.3$\pm$1.0\\ 
\textbf{B243}	 & 	B3 V	 & 	B8 V	 & 	$13500_{-1250}^{+1350}$ & $4.34_{-0.3}^{\uparrow}$ & $110_{\downarrow}^{+106}$ &$4.7_{-0.8}^{\uparrow}$ &$8.5_{-1.0}^{\uparrow}$ &$3.21_{-0.06}^{+0.07}$ &$7.5_{-0.8}^{+1.0}$ &$44_{-22}^{+21}$  & 20.2$\pm$1.5 \\ 
B253	 & 	B3V	 & 	B3-B5 III	 & 	$13000_{-1200}^{+2100}$ & $3.09_{\downarrow}^{+0.52}$ & $318_{\downarrow}^{\uparrow}$ &$4.1_{-0.2}^{+0.1}$ &$6.5_{-0.2}^{+0.1}$ &$3.21_{-0.06}^{+0.1}$ &$7.2_{-1.3}^{+1.1}$ &$3_{-1}^{+5}$ & 9.5$\pm$0.8 \\ 
\textbf{B268}	 & 	B2 V	 & 	B9-A0	 & 	$12250_{-1000}^{+850}$ & $3.99_{-0.38}^{\uparrow}$ & $36_{\downarrow}^{+126}$ &$4.6_{-0.8}^{\uparrow}$ &$8.1_{-1.0}^{\uparrow}$ &$3.24_{-0.05}^{+0.04}$ &$8.8_{-0.8}^{+1.2}$ &$31_{-15}^{+1}$ & 4.3$\pm$1.3 \\ 
\textbf{B275}	 & 	late-O\footnote{\label{P1:foot:SEDH97}From SED in \citetalias{1997ApJ...489..698H}}	 & 	B7 III	 &	$12950_{-650}^{+550}$ & $3.39_{-0.11}^{+0.06}$ & $88_{\downarrow}^{+68}$ &$3.8_{-0.8}^{+0.7}$ &$6.7_{-1.0}^{+0.8}$ &$3.37_{-0.03}^{+0.02}$ &$11.7_{-0.5}^{+0.67}$ &$8_{-2}^{+1}$ & -11.2$\pm$1.4\\ 
\underline{B289}	 & 	O9.5 V	 & 	O9.7V	 & 	$33800_{-2150}^{+2500}$ & $3.93_{\downarrow}^{+0.35}$ & $154_{-34}^{+38}$ &$3.7_{-0.5}^{+0.4}$ &$8.3_{-0.8}^{+0.6}$ &$4.84_{-0.07}^{+0.08}$ &$8.3_{-0.4}^{+0.4}$ &$19_{-5}^{+18}$ & -2.3$\pm$2.0 \\ 
B311	 & 	O9-B2 V	 & 	O8.5 Vz	 & 	$35950_{-1750}^{+2400}$ & $3.95_{\downarrow}^{+0.36}$ & $42_{-26}^{+12}$ &$3.3_{-0.3}^{+0.3}$ &$6.1_{-0.4}^{+0.4}$ &$4.79_{-0.06}^{+0.07}$ &$7.6_{-0.3}^{+0.2}$ &$14_{-4}^{+16}$ &4.2$\pm$0.4 \\ 
\textbf{B331}	 & 	B2 V\footnote{\label{P1:foot:H08}\citet{2008ApJ...686..310H}}	 & 	late-B	 & 	$13000$ & $-$ & $-$ &$4.6_{-0.5}^{+0.5}$ &$13.3_{-0.9}^{+0.9}$ &$4.10_{\downarrow}^{+0.37}$\footref{P1:foot:SED} &$21.8_{-7.2}^{+9.6}$ &$-$ & 13.9$\pm$1.5 \\ 
\textbf{B337}	 & 	mid-B\footref{P1:foot:SEDH97}	 & 	late-B	 & 	$13000$ & $-$ & $-$ &$3.7_{-0.7}^{+0.9}$ &$13.6_{-0.9}^{+0.9}$ &$3.38_{\downarrow}^{+0.02}$\footref{P1:foot:SED} &$8.7_{-3.3}^{+5.0}$ &$-$ & 7.5$\pm$0.6 \\ 
\hline
\vspace{-15pt}
\end{tabular}
\renewcommand{\footnoterule}{}
\end{minipage}
\label{P1:tab:Stellar_parameters}
\end{table*}

\subsection{Comparison of the radius estimated by two methods}

\begin{figure}
   \centering
  \includegraphics[width=\hsize]{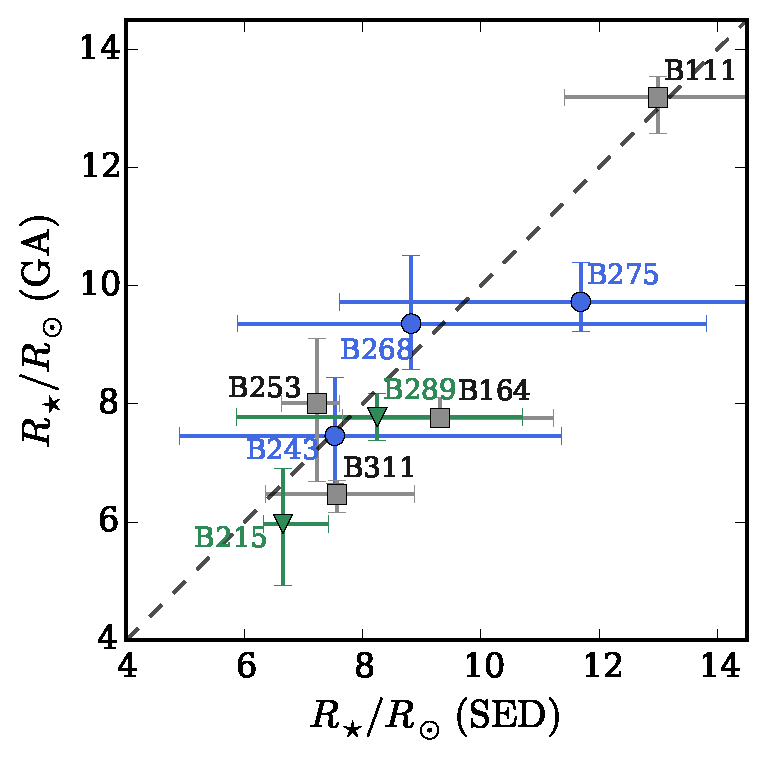}
      \caption{Comparison of the values for the stellar radius $R_{\star}$ obtained via SED fitting and GA analysis. The symbols and error bars correspond to the results obtained in Secs.~\ref{P1:SEDs} and \ref{P1:FASTWIND}; the dashed line represents the one to one correlation. The blue dots show the location of the targets for which we have found evidence of a gaseous disk, the green triangles show the stars that show IR excess longward of 2.5 $\mu$m and no emission lines in their spectra. The grey squares show the objects without disks. The values are consistent within the errors.}
         \label{P1:Fig:Rsed_vs_Rga}
   \end{figure}

The radius of the star is estimated in two different ways following the procedures described in the previous sections. In Fig.~\ref{P1:Fig:Rsed_vs_Rga} we show $R_{\star}$ obtained by fitting the SEDs to Castelli \& Kurucz models versus the one obtained via the genetic algorithm fitting. The diagonal line represents the one to one correlation and each symbol corresponds to one of our targets. B163, B331, and B337 were left out because for these stars it was not possible to identify any absorption lines, and therefore we did not include them in the GA fitting. The conclusion is that the two methods yield radii that are consistent within the errors.

\subsection{Hertzsprung-Russell diagram}
\label{P1:HRD}

\begin{figure}
   \centering
   \includegraphics[width=\hsize]{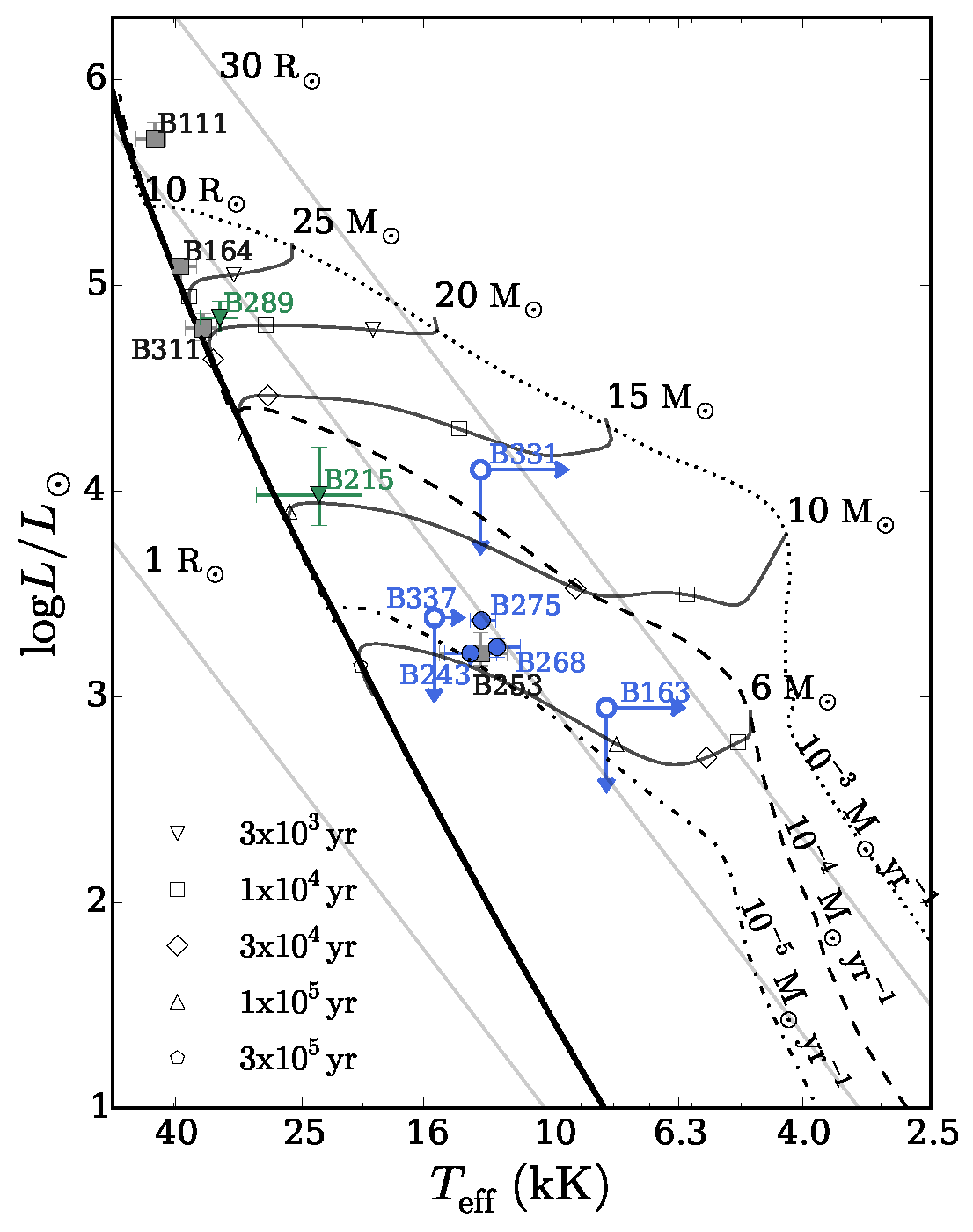}
      \caption{Hertzsprung-Russell diagram for the observed objects in M17. The blue dots show the location of the targets for which we have found evidence of a gaseous disk, the green triangles show the stars that show IR excess longward of 2.5 $\mu$m and no emission lines in their spectra. The grey squares show the objects without disk signatures. The stars for which the parameters are derived using \texttt{FASTWIND} are shown with filled symbols. The open symbols represent the stars for which $\log\,L/L_{\odot}$ and $T_{\rm eff}$ were derived from their spectral type. The ZAMS is represented by the thick black line and the solid lines correspond to PMS tracks from \citet{2009ApJ...691..823H}  with the ZAMS mass labeled and the lifetimes indicated as open symbols. The dotted, dashed and dashed-dotted lines are the birth lines for accretion rates of $10^{-3}$, $10^{-4}$ and  $10^{-5}$~$M_{\odot}$\,$\rm yr^{-1}$, respectively. The grey lines indicate radii of 1, 10 and 30~$R_{\odot}$.}
         \label{P1:Fig:HRD}
   \end{figure}

Figure~\ref{P1:Fig:HRD} shows the theoretical Hertzsprung-Russell diagram (HRD) based on the values of $T_{\rm eff}$ and $L$ obtained in the previous sections. We plotted the PMS tracks from \citet{2009ApJ...691..823H} with the ZAMS mass labeled and open symbols indicating lifetimes. We also present the birthline for accretion rates of $10^{-3}$, $10^{-4}$, and $10^{-5}$~$M_{\odot}yr^{-1}$. Assuming that the accretion is constant and that the Hosokawa tracks provide a correct description of the PMS evolution we conclude that 80\% of our sample must have experienced an on average high accretion rate (10$^{-3}$ - 10$^{-4}$~$M_{\odot}$~$\rm yr^{-1}$) and the remaining stars an accretion rate of at least 10$^{-5}$~$M_{\odot}$~$\rm yr^{-1}$.

All our confirmed PMS stars (blue circles) are located far away from the ZAMS and their positions can be compared with PMS tracks. The O stars (B111, B164, and B311; grey squares) and the stars without disk signatures (B289 and B215; green triangles) in the X-shooter spectra are located at or near the ZAMS. The position of the B-type star B253 is consistent with it being a PMS star, but its spectrum shows no signatures for the presence of a circumstellar disk.
 
\section{Evidence for the presence of circumstellar disks}
\label{P1:NIR}

In this section we present an overview of the emission-line features thought to be produced by the circumstellar disks. We discuss the nature of the disks of the PMS objects identified in the previous sections and measure the disk rotational velocity, $V_{\rm disk}$, using several hydrogen lines. In some of the objects we identify CO bandhead emission. The infrared excess observed in the SEDs (Sect.~\ref{P1:SEDs}) provides information on the dust component of the disk.

\begin{figure*}
   \centering
  \includegraphics[width=\hsize]{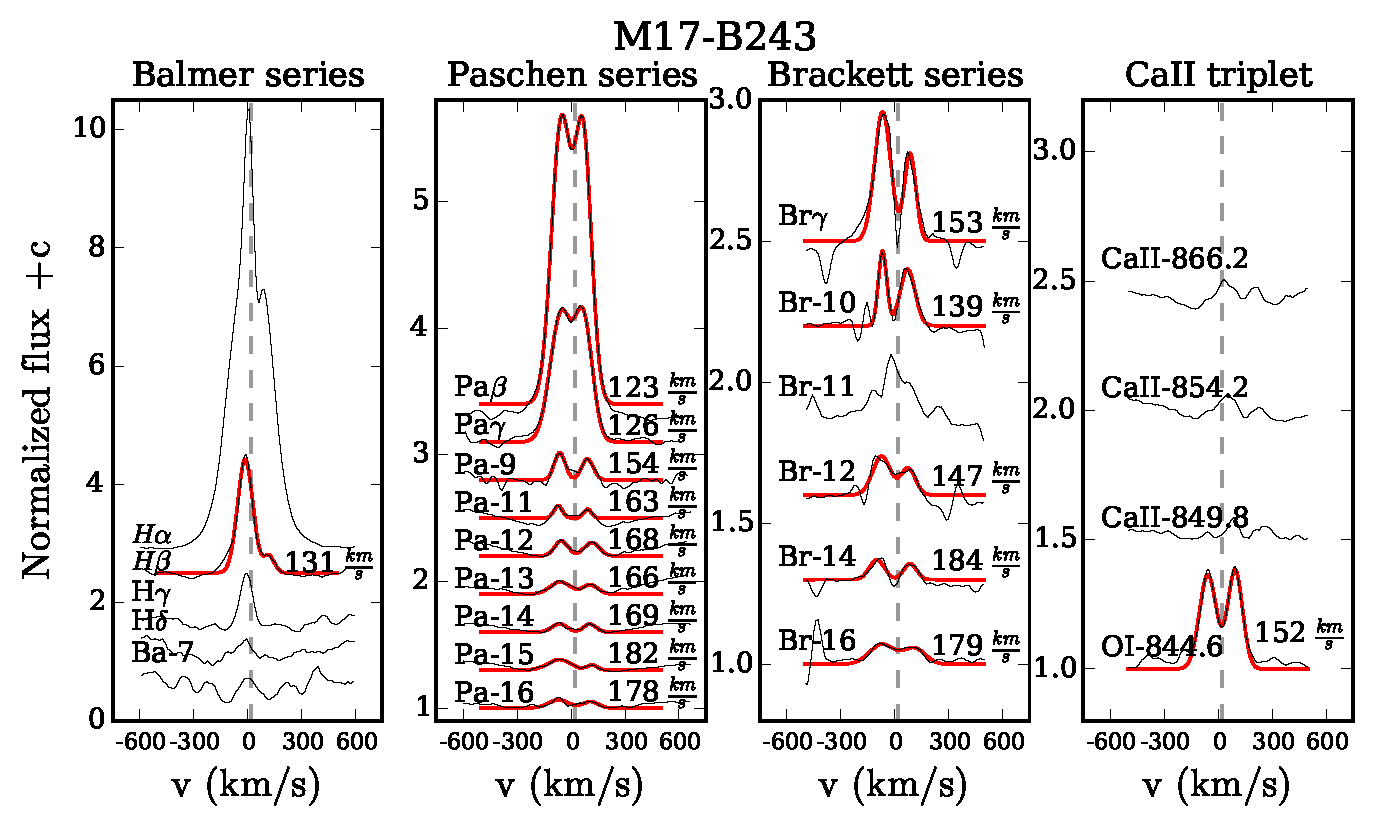}
      \caption{Double-peaked emission lines in M17-B243. These profiles indicate the presence of a circumstellar rotating disk. The X-shooter spectrum is shown with the solid black lines and the two Gaussian functions fitted to the profiles are shown in red. The lines are plotted in the heliocentric frame and the dashed grey line represents the radial velocity of the star. Each line is labeled in the left part of the plots and the peak to peak separation is displayed at the right side of the profiles.}
         \label{P1:Fig:doublePeakedVel_B243}
   \end{figure*}

\begin{figure*}
   \centering
  \includegraphics[width=\hsize]{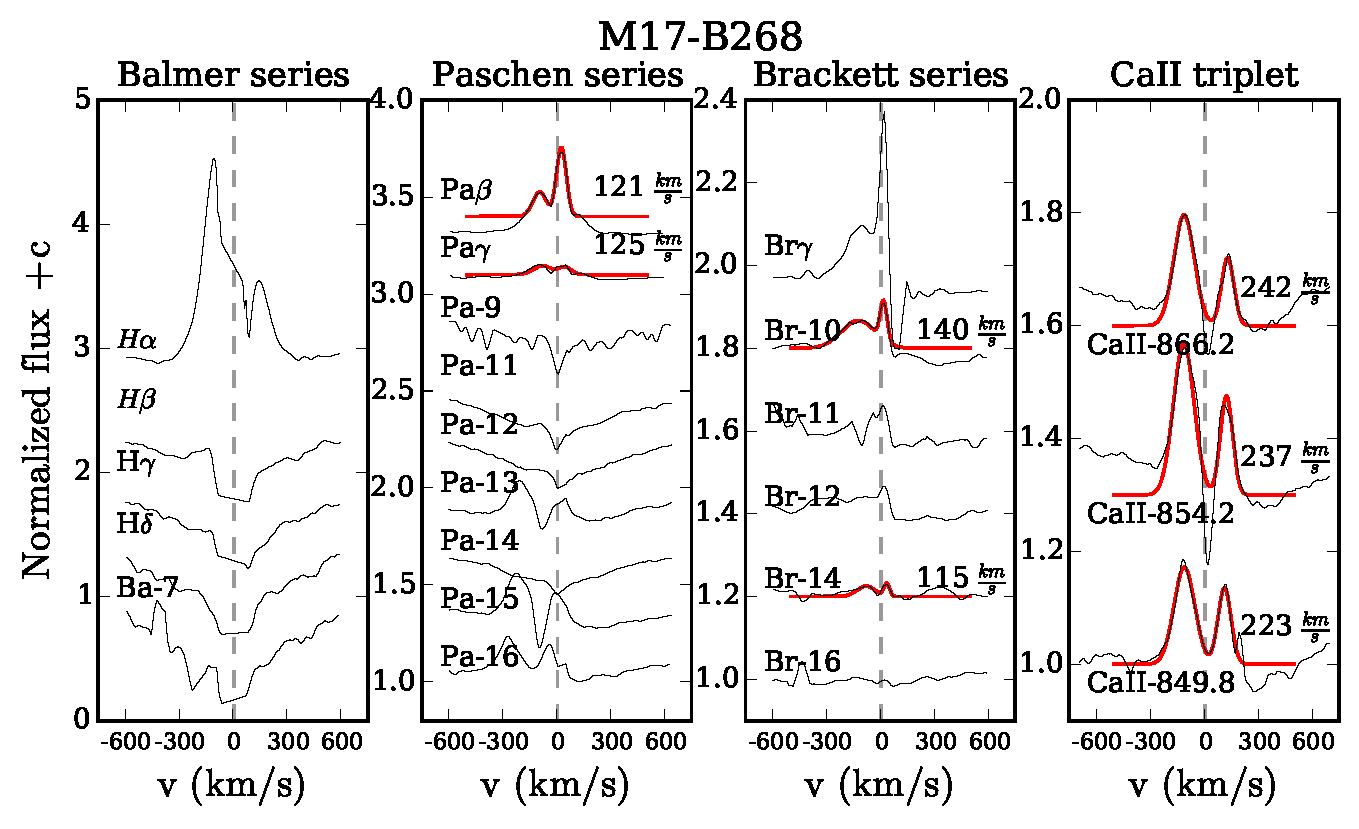}
      \caption{Same as Fig.~\ref{P1:Fig:doublePeakedVel_B243} for M17-B268.}
         \label{P1:Fig:doublePeakedVel_B268}
   \end{figure*}

\begin{figure*}
   \centering
  \includegraphics[width=\hsize]{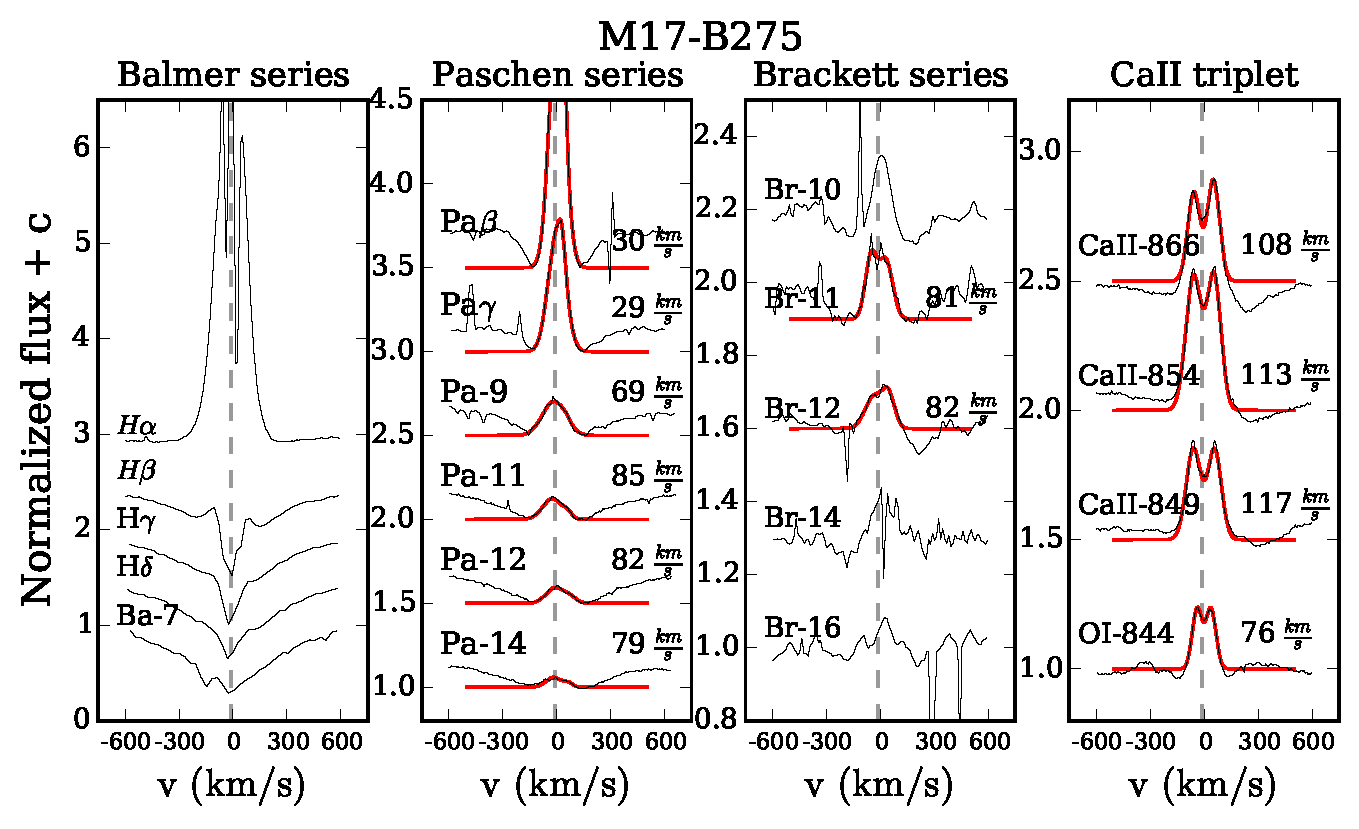}
      \caption{Same as Fig.~\ref{P1:Fig:doublePeakedVel_B243} for M17-B275.}
         \label{P1:Fig:doublePeakedVel_B275}
   \end{figure*}

\begin{figure}
   \centering
  \includegraphics[width=\hsize]{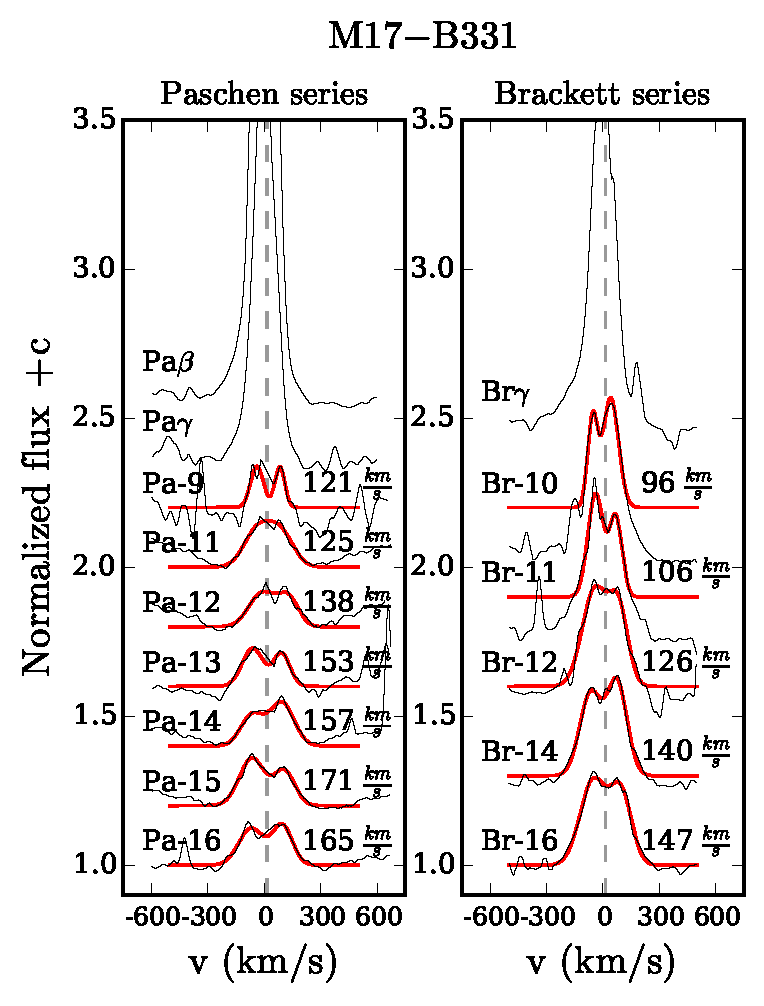}
      \caption{Same as Fig.~\ref{P1:Fig:doublePeakedVel_B243} for M17-B311.}
         \label{P1:Fig:doublePeakedVel_B331}
   \end{figure}

\begin{figure}
   \centering
  \includegraphics[width=5cm]{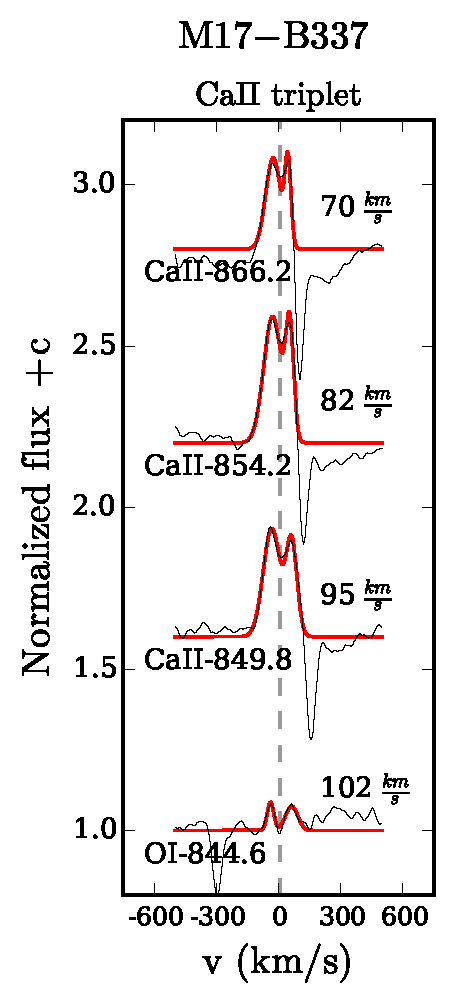}
      \caption{Same as Fig.~\ref{P1:Fig:doublePeakedVel_B243} for M17-B337.}
         \label{P1:Fig:doublePeakedVel_B337}
   \end{figure}

We observe several double-peaked emission lines and/or CO bandhead emission in the visual to near-infrared spectrum in six of our targets: B163, B243, B268, B275, B331, and B337. A selection of double-peaked lines along the X-shooter wavelength range is shown in Figs. \ref{P1:Fig:doublePeakedVel_B243} to \ref{P1:Fig:doublePeakedVel_B337}.  The fact that we see double-peaked emission is indicative of a rotating circumstellar disk. For all of these objects we observe an infrared excess in their SEDs (Sect.~\ref{P1:SEDs}). The atypical morphologies observed in the Balmer and Brackett series of B243 and B268 might be an indication of active accretion; nevertheless, further higher resolution observations are needed to confirm this scenario.

\subsection{Velocity structure of the gaseous disks}

To measure the characteristic projected rotational velocity properties of the gaseous component of the circumstellar disks, we selected a sample of double-peaked emission lines for each of the sources and fitted two Gaussian functions to measure the peak to peak separation. The projected rotational velocity measured in this way corresponds to half of the peak to peak separation. The Gaussian functions fitted to the lines are shown with the solid red lines in Figs. \ref{P1:Fig:doublePeakedVel_B243} to \ref{P1:Fig:doublePeakedVel_B337} and the results from these measurements are listed in Tab.~\ref{P1:tab:peak-separation}. The Balmer, Paschen, and Brackett series and \ion{Ca}{ii} triplet lines are plotted in each of the columns (from left to right) and lines are labeled in the left part of the plot. The measured peak to peak separation is indicated to the right.

\begin{table}
\centering
\caption{\normalsize{Projected disk rotational velocity measured from half the peak to peak separation of the lines originating in the circumstellar disk.}}
\renewcommand{\arraystretch}{1.4}
\setlength{\tabcolsep}{2pt}
\begin{tabular}{lcccccc}
\hline
\hline
 & \multicolumn{5}{c}{$v_{\rm disk}$~[km/s]} \\
Line & B243 & B268 & B275 & B331 & B337  \\ 
 \hline 
OI-844.6 & 76 & - & 38 & - & - \\ 
CaII-849.8 & - & 112 & 59 & - & 48 \\ 
Pa-16 & 86 & - & - & 83 & - \\ 
CaII-854.2 & - & 119 & 57 & - & 41 \\ 
Pa-15 & 91 & - & - & 85 & - \\ 
Pa-14 & 86 & - & 40 & 81 & - \\ 
CaII-866.2 & - & 122 & 54 & - & 35 \\ 
Pa-13 & 83 & - & - & 80 & - \\ 
Pa-12 & 84 & - & 43 & - & - \\ 
Pa-11 & 79 & - & 43 & 64 & - \\ 
Pa-9 & 78 & - & 45 & 63 & - \\ 
Pa$\gamma$ & 63 & 63 & 29 & - & - \\ 
Pa$\beta$ & 62 & 58 & 28 & - & - \\ 
Br-16 & 90 & - & - & 74 & - \\ 
Br-14 & 92 & 58 & - & 70 & - \\ 
Br-12 & 73 & - & 44 & 64 & - \\ 
Br-11 & - & - & 40 & 53 & - \\ 
Br-10 & 70 & 72 & - & 48 & - \\ 
Br$\gamma$ & 77 & 62 & - & - & - \\ 
\hline
\vspace{-15pt}
\end{tabular}
\renewcommand{\footnoterule}{}
\label{P1:tab:peak-separation}
\end{table}

\begin{figure}
  \includegraphics[width=\hsize]{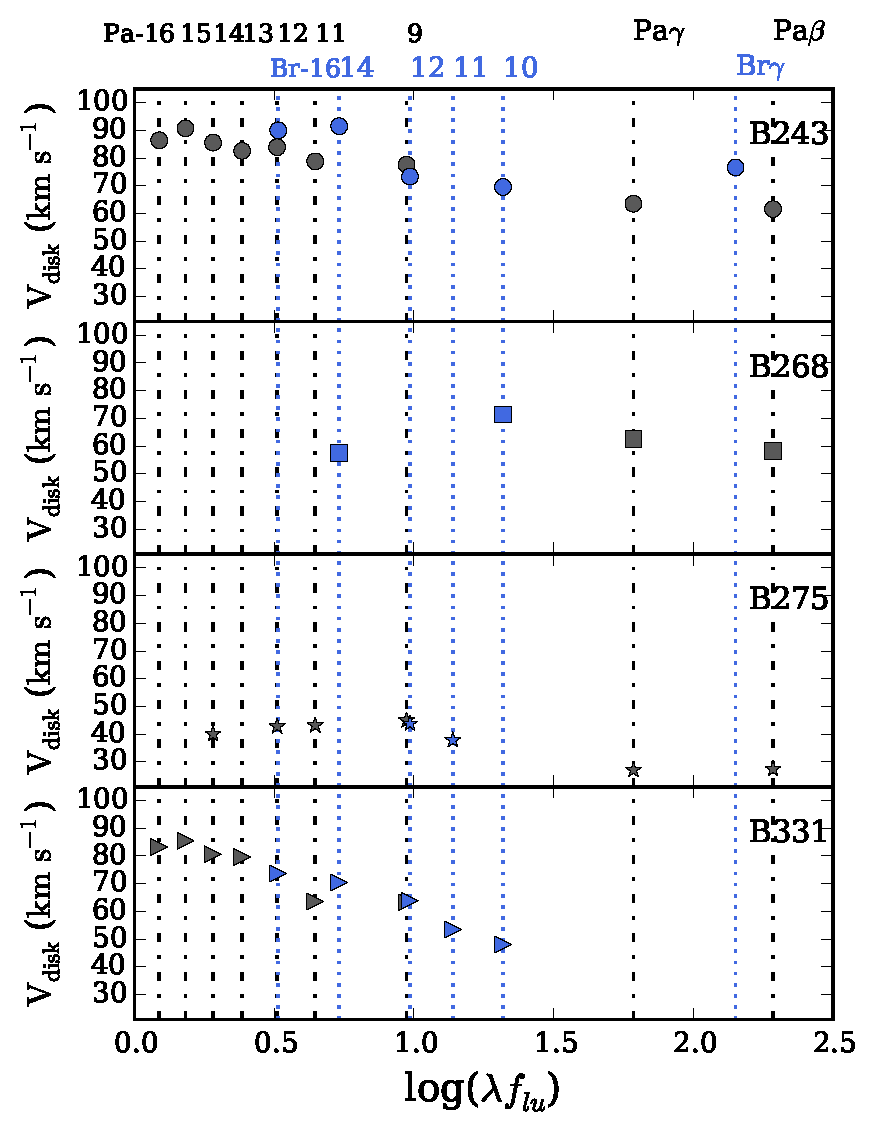}
      \caption{The projected disk rotational velocity, $\rm V_{disk}$, against oscillator strength for double-peaked lines along the wavelength coverage of X-shooter for the stars indicated in the upper-right corner. The Paschen series with dashed-dotted lines (black) and the Brackett series with dotted lines (blue).}
         \label{P1:Fig:vsini_vs_lambdaoscstr}
   \end{figure}

Figure~\ref{P1:Fig:vsini_vs_lambdaoscstr} shows the disk projected velocities measured from each of the hydrogen lines against $\rm \log{ \lambda f_{lu}}$, where $f_{lu}$ is the oscillator strength. We use  $\rm \log{ \lambda f_{lu}}$ as a measure of the relative strength of a given line within a line series. The oscillator strengths for the hydrogen lines were obtained from \citet{1968ApJS...17..445G}. For B163 and B337 we do not have sufficient velocity measurements to draw any conclusions. For four out of the six gaseous disks detected (B243, B268, B275, and B331) we see a clear trend of the projected disk rotational velocity from the hydrogen recombination lines with line strength. This suggests a structured velocity profile of the gaseous disks, in qualitative agreement with the prediction that the high excitation lines form in the inner region (dense and hot) of the disk while the low excitation lines form over a larger area (and on average more slowly rotating part) of the disk.

B268 and B275 also exhibit \ion{Ca}{ii} triplet and \ion{O}{i} double-peaked emission lines. The velocities measured from these lines are higher than those of the hydrogen lines. This indicates that these lines are formed closer to the star than the hydrogen lines.

Assuming that the disks have a Keplerian velocity structure and adopting the mass of the central object, it is possible to roughly calculate the distance from the star at which the lines are formed. This results in a range from a few hundred to one thousand~$R_{\odot}$ (tens to a few hundreds $R_{\star}$).

\subsection{CO bandhead emission}

CO overtone emission is produced in high-density (10$^{10}$ -- 10$^{11}$~cm$^{-3}$) and high-temperature (2500 -- 5000~K) environments \citep{1995ApJS..101..309C}. CO is easily dissociated, and therefore must be shielded from the strong UV radiation coming from the star. These conditions are expected in the inner regions of (accretion) disks, which makes the CO bandheads an ideal tool to trace the disk structure around mYSOs \citep[e.g. ][]{2006A&A...455..561B, 2013MNRAS.429.2960I}. The shape of the CO lines can be modelled by a circumstellar disk in Keplerian rotation \citep{2004A&A...427L..13B, 2004ApJ...617.1167B, 2010MNRAS.402.1504D, 2010MNRAS.408.1840W, 2013MNRAS.429.2960I}. The blue shoulder in the bandheads is a measure of the inclination of the disk: an extended blue shoulder indicates a high inclination angle (i.e., near "edge-on" view). 

We detect CO-bandhead emission in B163, B243, B268, B275, and B331 (see Fig.~\ref{P1:fig:NIR}).  Although CO-bandhead emission is rare, it is also seen in some B[e] supergiants \citep{1998A&A...340..117L}. However, these evolved B[e] stars show numerous forbidden emission lines in their optical spectra, while we only observe the well known [\ion{O}{i}]\,6300 in some of the sources. Hence, the observed spectra support our hypothesis that the observed disks are related with the accretion process in contrast to a possible origin such as in (more evolved) B[e] stars \cite[see][]{2004A&A...427L..13B}. Modelling of the bandheads and the double-peaked hydrogen emission lines will be the subject of a forthcoming paper.

 \section{Discussion}
 \label{P1:Discussion}

\subsection{Age distribution of the massive PMS population in M17}
\label{P1:age}

Age estimates of M17 show that its stellar population is not older than 1 Myr. There is no evidence for a supernova explosion in that area \citep{2007ApJS..169..353B}, consistent with such a young age. We estimated the age of the main-sequence stars by comparing their position in the HRD with Milky Way evolutionary tracks from \citet{2011A&A...530A.115B}. We compare the position of our stars with evolutionary tracks that account for the effects of rotation considering a flat distribution of spin rates starting at the measured $v\sin{i}$. To do so, we used the bayesian tool \texttt{BONNSAI}\footnote{The \texttt{BONNSAI} web-service is available at \url{www.astro.uni-bonn.de/stars/bonnsai}.} from \citet{2014A&A...570A..66S}. Within the uncertainties of the models, the ages obtained for B111, B164, and B311 are 1.60, 0.98, and 0.82~Myr, in fair agreement with the age estimates for the region (Tab.~\ref{P1:tab:ages_bonnsai_PMS}). As a way of testing our classification of the PMS stars, we estimated their ages using main sequence tracks through \texttt{BONNSAI}. We obtained main sequence ages ranging from 10 to 90~Myr, which would be inconsistent with M17 being a young region. Assuming that our classified PMS stars are part of M17, we conclude that these stars cannot be post-, but must be pre-main sequence objects.

We estimated the age of the PMS stars by comparing their effective temperatures and luminosities with PMS tracks \citep{2009ApJ...691..823H}. For these stars we find an age span from tens of thousands of years to a few hundred thousand years. 

If we compare the position of the (presumable) main-sequence stars B215, B253, and B289 to isochrones, we obtain age estimates of $\sim$9, $\sim$50, and $\sim$4~Myr, respectively. B253 does not present emission lines nor IR excess in its SED, and B215 and B289 do not present emission lines in the X-shooter spectrum but have IR excess longward of 2.3~$\mu$m. If these sources are PMS stars this suggests that their circumstellar disks have already (at least partially) disappeared (see also Sect. \ref{P1:ext}).

\begin{table}
\centering
\caption{\normalsize{Age of our targets estimated by comparing their position in the HRD with PMS tracks from \citet{2009ApJ...691..823H} (third column) and main sequence tracks from \citet{2011A&A...530A.115B} (fourth column).}}
\renewcommand{\arraystretch}{1.4}
\setlength{\tabcolsep}{4pt}
\begin{tabular}{lccc}
\hline
\hline
star & PMS & PMS  & MS \\
& track & lifetime & lifetime \\
& [$M_{\odot}$] & [Myr] & [Myr] \\
 \hline 
B111	&	25	&	$-$ 	&	$0-2$ \\ 
B163	&	6 	&	0.14	&	$-$	 	\\
B164	&	25	&	0.01	&	$0-2$  \\
B215	&	10	&	0.03 	&	$-$	 	\\
B243	&	6 	&	0.20	&	$-$	 	\\
B253	&	6 	&	0.20	&	$-$	 	\\
B268	&	6 	&	0.20	&	$-$	 	\\
B275	&	8 	&	0.04	&	$-$	 	\\
B289	&	20	&	0.02	&	$-$	 	\\
B311	&	20	&	0.02	&	$0-2$  \\ 
B331	&	12	&	0.02	&	$-$	 	\\
B337	&	8	&	0.04	&	$-$	 	\\
\hline
\vspace{-15pt}
\end{tabular}
\renewcommand{\footnoterule}{}
\label{P1:tab:ages_bonnsai_PMS}
\end{table}

Given the size of our sample and the uncertainties in the models we cannot draw a firm conclusion about the age of M17 nor about the possible presence of two distinct populations in this region \citep{2009ApJ...696.1278P}. Nevertheless we observe a trend of age with luminosity: the less luminous objects are further away from the ZAMS than their more luminous counterparts. This is in line with the idea \citep[][]{2003ApJ...585..850M, 2007ARA&A..45..481Z, 2011MNRAS.416..972D} that more massive stars form faster and therefore spend less time on the PMS tracks than lower-mass stars.

\subsection{Extinction towards the PMS stars in M17}
\label{P1:ext}

Table~\ref{P1:tab:Stellar_parameters} lists the extinction properties of our targets. We find that the extinction in the $V$-band varies from $\sim$5 to $\sim$14 mag and that the total to selective extinction is $3.3 < R_{V} < 5$. The two sources with the highest extinction are situated near or in the irradiated molecular cloud (see Fig.\ref{P1:fig:M17rgb}), but there is no general trend with location in the H\,~{\sc ii} region. The sources with higher $A_V$ tend to have larger values of $R_V$, although the correlation is weak. The overall conclusion is that the extinction is quite patchy, with substantial variation on a spatial scale of 50 arcsec (corresponding to a geometrical scale of 0.5~pc at the distance of M17), similar as to the findings of \citetalias{1997ApJ...489..698H}. A dust disk local to the star, for example, may dominate the line-of-sight extinction towards the sources \citep[see][]{2013A&A...558A.102E}. Studies of individual sources therefore should not rely on average properties of the region, but should be based on a detailed investigation of the extinction in the line of sight.

\subsection{Presence of circumstellar disks around massive PMS stars}

We detect signatures of circumstellar disks in six of our sources: B163, B243, B268, B275, B331, and B337. The full sample includes two O-type stars, B111 and B164, that are not classified as potential YSO sources by \citetalias{1997ApJ...489..698H}, but as main sequence objects, and do not reveal disk signatures. Two stars, B215 (IRS15) and B289, do not show evidence for gaseous disk material in their spectral lines but do feature excess infrared continuum emission indicative of dust in the circumstellar environment. We thus find that 60\% of our massive YSO candidates show clear evidence for circumstellar disks in their spectrum and another 20\% likely also feature disks based on NIR excess emission. 

In Fig.~\ref{P1:Fig:HRD} we have labeled the sources with a gaseous disk detectable in the X-shooter spectrum with a blue dot, the ones with only NIR excess in the SED with a green triangle and the O and B stars with a grey square. The stars with gaseous disks are located further away from the ZAMS than the other sources. However, we cannot link this in a straightforward way to an evolutionary effect. Having noted this, and taking into account that our sources with non-detectable disks are younger than $10^5$~yrs, we can conclude that massive stars up to $\sim$20~$M_{\odot}$ retain disks up to less than $10^{5}$ years upon arrival on the ZAMS. We identify the presence of [\ion{O}{i}]\,6330 emission in three sources (B215, B243, and B275). In  Herbig Ae/Be this line is mostly formed in stars with a flaring disk \citep{2008A&A...485..487V}.  The spectral profiles of some lines in B268 and B243 might be an indication that this star is actively accreting. For the other sources, as we do not find strong evidence for infall or the presence of jets, we do not know whether these sources are actively accreting or that the observed disks are structures remnant of the formation process.

\subsection{Spin properties of the massive PMS stars in M17}

The projected rotational velocity ($v\sin{i}$) distribution for 216 O stars in 30 Dor has been published by \citet{2013A&A...560A..29R}. Their distribution shows a two-component structure consisting of a peak at 80~$\rm km$~$\rm s^{-1}$ with 80\% of the stars having $0< v\sin{i} < 300$~$\rm km$~$\rm s^{-1}$ and a high velocity tail (containing 20\% of the stars) extending up to 600~$\rm km$~$\rm s^{-1}$. \citet{2013A&A...550A.109D} studied 300 stars spanning spectral types from O9.5 to B3 in 30\,Dor; they find a bimodal distribution with 25\% of the stars with $0 < v\sin{i} < 100$~$\rm km$~$\rm s^{-1}$ and a high velocity tail between $200 < v\sin{i} < 350$~$\rm km$~$\rm s^{-1}$. 
They estimated $v\sin{i}$ using a Fourier transform method, which allows them to separate the rotational broadening from other broadening mechanisms. \citet{2009ApJ...700..844P} measured $v\sin{i}$ for 97 OB stars in the Milky Way. They find that 80\% of their sample rotate slower than 200~$\rm km$~$\rm s^{-1}$, and that the remaining 20\% has $200 < v\sin{i} < 400$~$\rm km$~$\rm s^{-1}$. Tab.~\ref{P1:tab:Stellar_parameters} shows the $v\sin{i}$ measured for the stars in our sample. We find that around 30\% of our sample is rotating relatively fast (around 200\,$\rm km$~$\rm s^{-1}$).

\begin{figure}
  \includegraphics[width=\hsize]{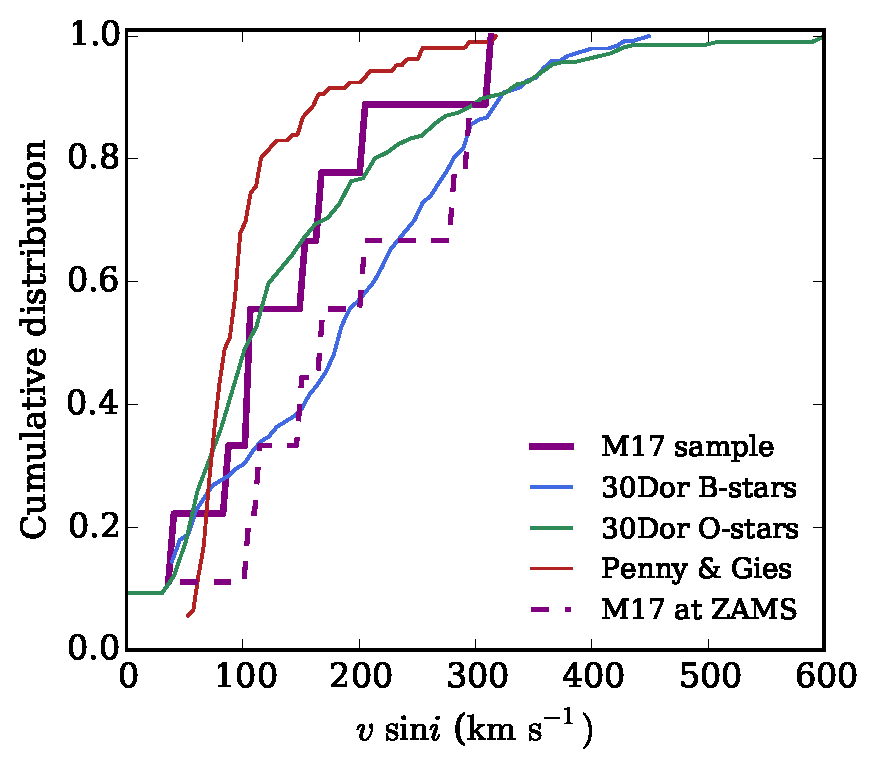}
      \caption{Comparison of the cumulative distribution of projected rotational velocities of our sample (purple, 9 stars); \citet{2013A&A...550A.109D} (blue, 251 stars); \citet{2013A&A...560A..29R} (green, 216 stars); and \citet{2009ApJ...700..844P} (red, 106 stars). The dashed line indicates the distribution of projected rotational velocities of our stars upon arrival at the ZAMS assuming rigid rotation (analogous to homologous contraction) and angular momentum conservation.}
         \label{P1:Fig:cumulative_dist}
   \end{figure}

We show the cumulative $v\sin{i}$ distribution functions of these works together with our results in Fig.~\ref{P1:Fig:cumulative_dist}. To quantitatively compare these distributions we performed a Kuiper test, which allows to test the null hypothesis that two observed distributions are drawn from the same parent distribution. The significance level of the Kuiper statistic, $p_K$, is a percentage that indicates how similar the compared distributions are. Small values of $p_K$ show that our cumulative distribution is significantly different from the one it is compared to. The $p_K$ values obtained from comparing our distribution with the ones of \citet{2009ApJ...700..844P}, \citet{2013A&A...560A..29R}, and \citet{2013A&A...550A.109D} are $p_K=13\%$, 99\%, and 18\%, respectively. As these values are not lower than 10\%, they do not allow us to reject the null hypothesis. 

Upon arrival on the ZAMS our sample will span a roughly similar range of masses as the B-star sample in 30\,Dor \citep{2013A&A...550A.109D}. Therefore, it is most appropriate to compare our findings to this sample. According to the models by \citet{2011A&A...530A.115B} and \citet{2012A&A...537A.146E} the $v\sin{i}$ distribution of stars in our mass range will not change significantly during the first few Myrs of evolution. \citet{2013A&A...550A.109D} point out that macroturbulent motions only need to be taken into account in cases of relatively slow spinning stars ($v\sin{i}<80$~km\,$\rm s^{-1}$). We note that \citet{2013A&A...550A.109D} discuss the possibility of their presumably single star sample to be polluted by (relatively rapidly spinning) post-interaction binaries.
We estimated the $v\sin{i}$ that our PMS stars will have upon arrival on the ZAMS using the ZAMS radii corresponding to the end of the PMS tracks from \citet{2009ApJ...691..823H}. We applied angular momentum conservation and assumed that the stars are rigidly rotating (which is analogous to homologous contraction). When comparing this $v\sin{i}$ distribution (dashed line in Fig.~\ref{P1:Fig:cumulative_dist}) with the B-star sample in 30\,Dor we find that $p_k = 87\%$. 
Assuming that our stars represent a progenitor population of the LMC B-star sample studied by \cite{2013A&A...550A.109D} (neglecting metallicity effects and pollution from post-interaction systems in the B stars in 30\,Dor sample) we find that the contraction of the PMS stars during the main sequence is consistent with being homologous. Of course, it is premature to firmly state this given the small sample size and the caveats mentioned above.

\section{Conclusions}
\label{P1:Conclusions}

We performed VLT/X-shooter observations of young OB stars in M17. We classified and modelled the photospheric spectra using \texttt{FASTWIND} in order to derive their stellar parameters. We identified the presence of gaseous and dusty disks in some of them based on the emission lines in the spectrum and on the IR excess observed in the spectral energy distribution. We confirm the PMS nature of six objects in this region and conclude that they are on their way of becoming B main sequence stars with masses ranging from 6 to 20~$M_{\odot}$. This constitutes a unique sample of PMS stars that allows us to test theoretical star formation models. Our findings can be summarised as follows:

\begin{itemize}
\item We confirm the PMS nature of six of the mYSO candidates presented by \citetalias{1997ApJ...489..698H}. We conclude that most of our objects must have experienced, on average, high accretion rates.

\item The age of the PMS objects has been obtained by comparing their position in the HRD with pre-main sequence tracks of \citet{2009ApJ...691..823H}, whereas the age of the OB stars was estimated using the tool \texttt{BONNSAI} \citep{2014A&A...570A..66S} and comparing to tracks of \citet{2011A&A...530A.115B}. For the O stars we obtained ages younger than 2~Myr. The pre-main sequence stars have estimated PMS lifetimes of a few hundred thousand years. Given the uncertainty in the age and the fact that we have only a few stars, we cannot conclude that the PMS population corresponds to a second generation of star formation in M17 but found no indication in favour either.

\item We measured the visual and total to selective extinction towards our objects by fitting Castelli \& Kurucz models corresponding to the spectral types. We confirm that the extinction towards M17 is highly variable, as usually observed in star forming regions. We point out that a dust disk local to the star may dominate the line-of-sight extinction towards the sources. Based on the NIR excess (> 2~$\mu$m) observed in the spectral energy distributions we found dusty disks in eight of our targets.

\item Via (double-peaked) emission lines we found evidence for gaseous circumstellar disks in six of our targets, all of which also include a dust component. We measured the projected rotational velocity of the disks from each of the double-peaked lines and found a structured velocity profile among the hydrogen recombination lines for four out of the six disks. For two out of the six disks we were able to identify \ion{Ca}{ii} triplet and \ion{O}{i} double-peaked emission lines. The velocities measured from these lines are larger than from the ones measured from hydrogen suggesting that they are formed closer to the star than the hydrogen lines.  

\item We measured the projected rotational velocities, $v\sin{i}$, of our stars. About 30\% of our sample rotate at around 200~$\rm km$~$\rm s^{-1}$ or faster. The PMS objects are expected to contract and therefore have spun up upon arrival on the main-sequence. Assuming homologous contraction and the absence of processes causing angular momentum loss in their remaining PMS evolution, the $v\sin{i}$ distribution of our sample appears consistent with that of the B stars in 30\,Dor once they ignite hydrogen. We note that the contraction in PMS stars is not well understood. Two of the objects will have a $v\sin{i} >300$~$\rm km\,s^{-1}$ upon arrival on the MS.

\end{itemize}

With this unique, though still small sample we show the potential for constraining models of star formation. A larger sample is needed in order to robustly assess the validity of different theories.

\begin{acknowledgements}
We thank the anonymous referee for for carefully reading the manuscript and many helpful and insightful comments and suggestions. This research has made use of the SIMBAD database, operated at CDS, Strasbourg, France. This research made use of Astropy, a community-developed core Python package for Astronomy (Astropy Collaboration, 2013). MCRT is funded by the \emph{Nederlandse Onderzoekschool Voor Astronomie} (NOVA).
\end{acknowledgements}

%-------------------------------------------------------------------

\bibliography{/Users/macla/Dropbox/bibmaster/references}

\onecolumn
\begin{appendix}

\Online

 \section{Full X-shooter spectra.}
 \label{P1:appendix_spec}

Normalised VLT/X-shooter spectra of the objects studied in this paper (Table~1), combining the UVB, VIS and NIR arm. Line identifications are indicated above the spectrum. Note the presence of telluric absorption features centred at 690, 720, 820, 920, 1400, 2010 and 2060 nm. The attenuated region between the H- and the K band (1830 - 1980 nm) has been omitted. The blue part of the spectrum for B163, B331, and B337 is not visible due to severe interstellar extinction. We have included the spectra of B111, B275, and B311 as an example, the version of the appendix for the full sample will be available on A\&A.

\clearpage
\newpage

  \begin{figure}[h]
  \includegraphics[width=\hsize]{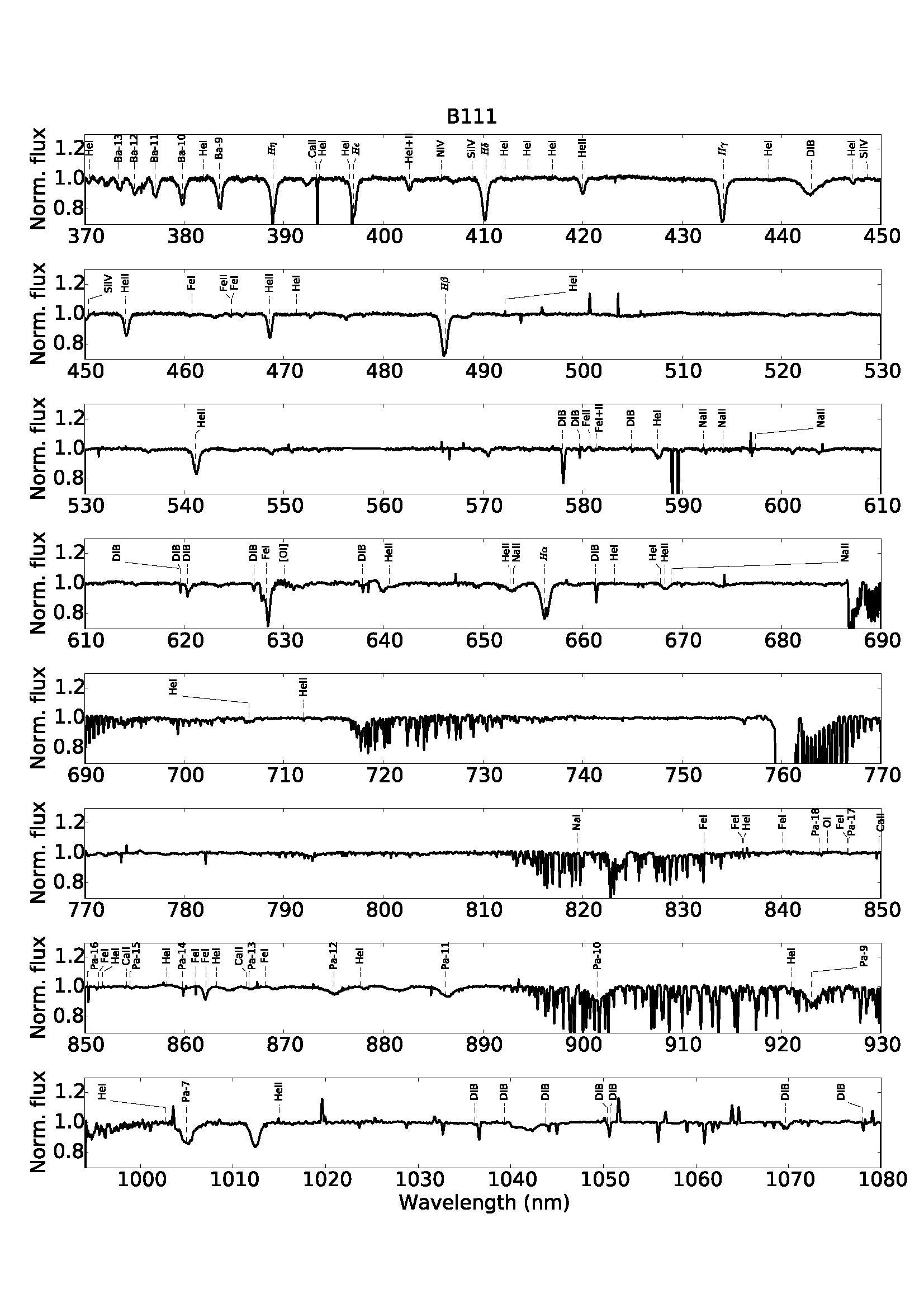}
   \end{figure}
  \begin{figure}[h]
  \includegraphics[width=\hsize]{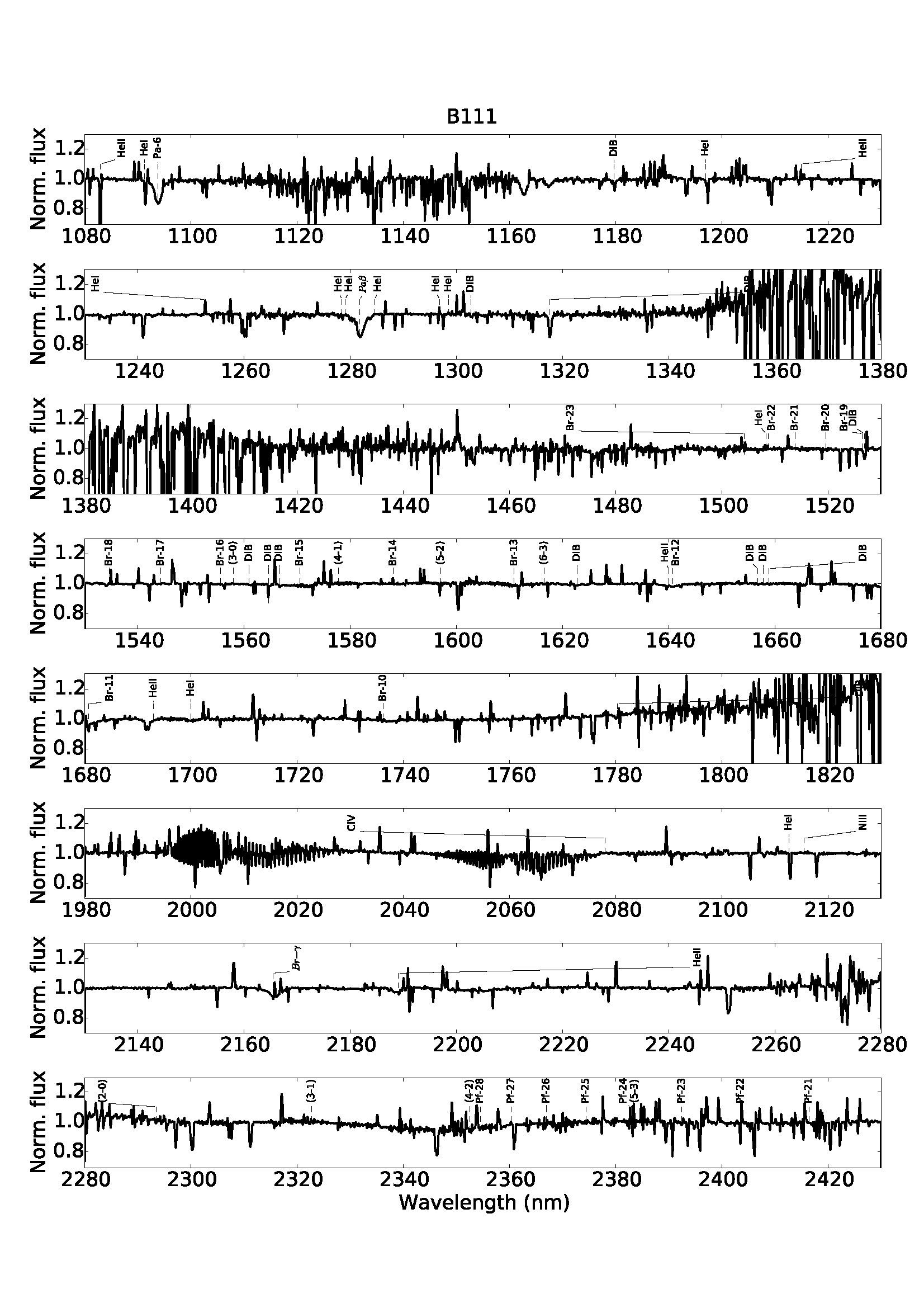}
   \end{figure}
  \begin{figure}[h]
  \includegraphics[width=\hsize]{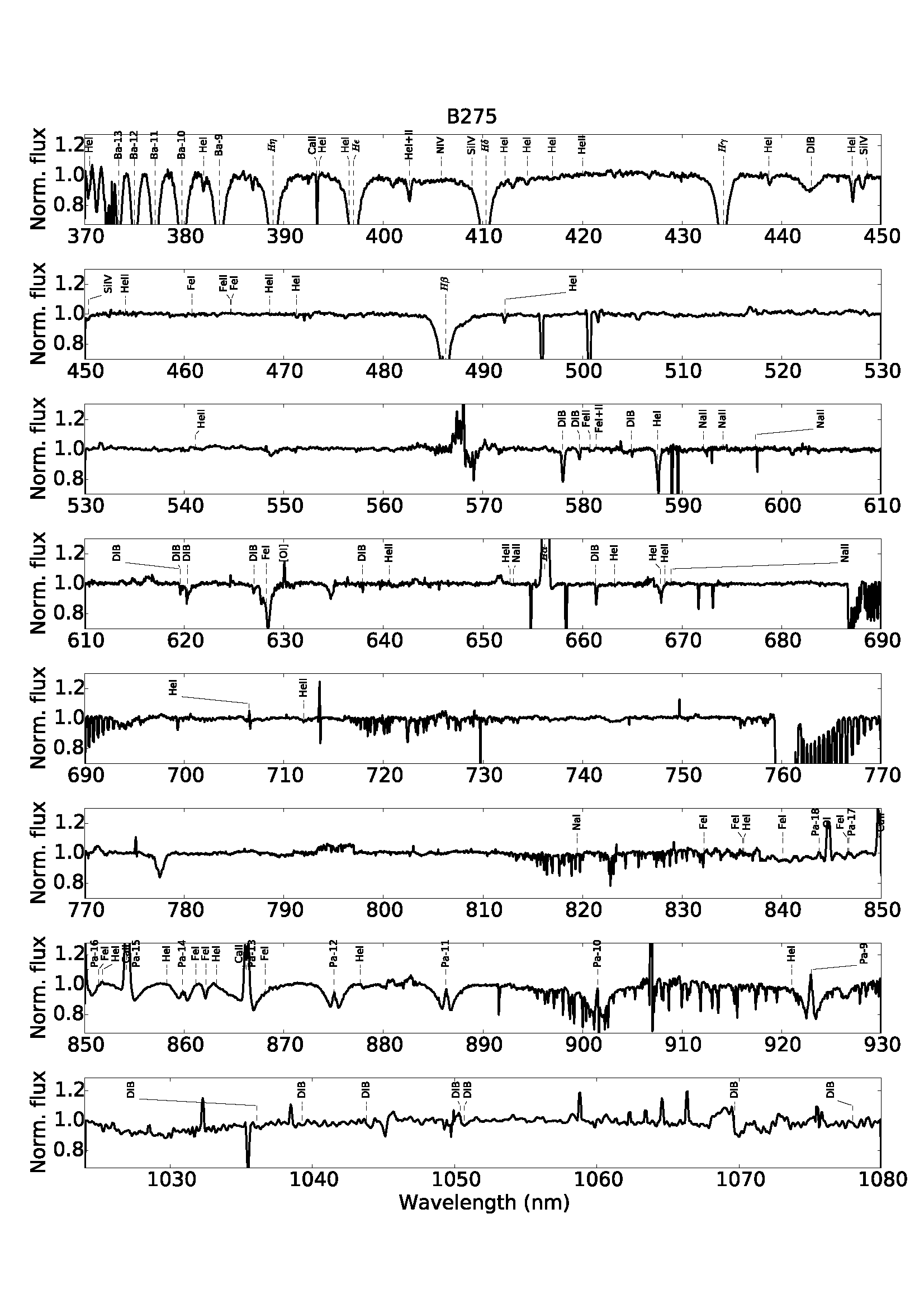}
   \end{figure}
  \begin{figure}[h]
  \includegraphics[width=\hsize]{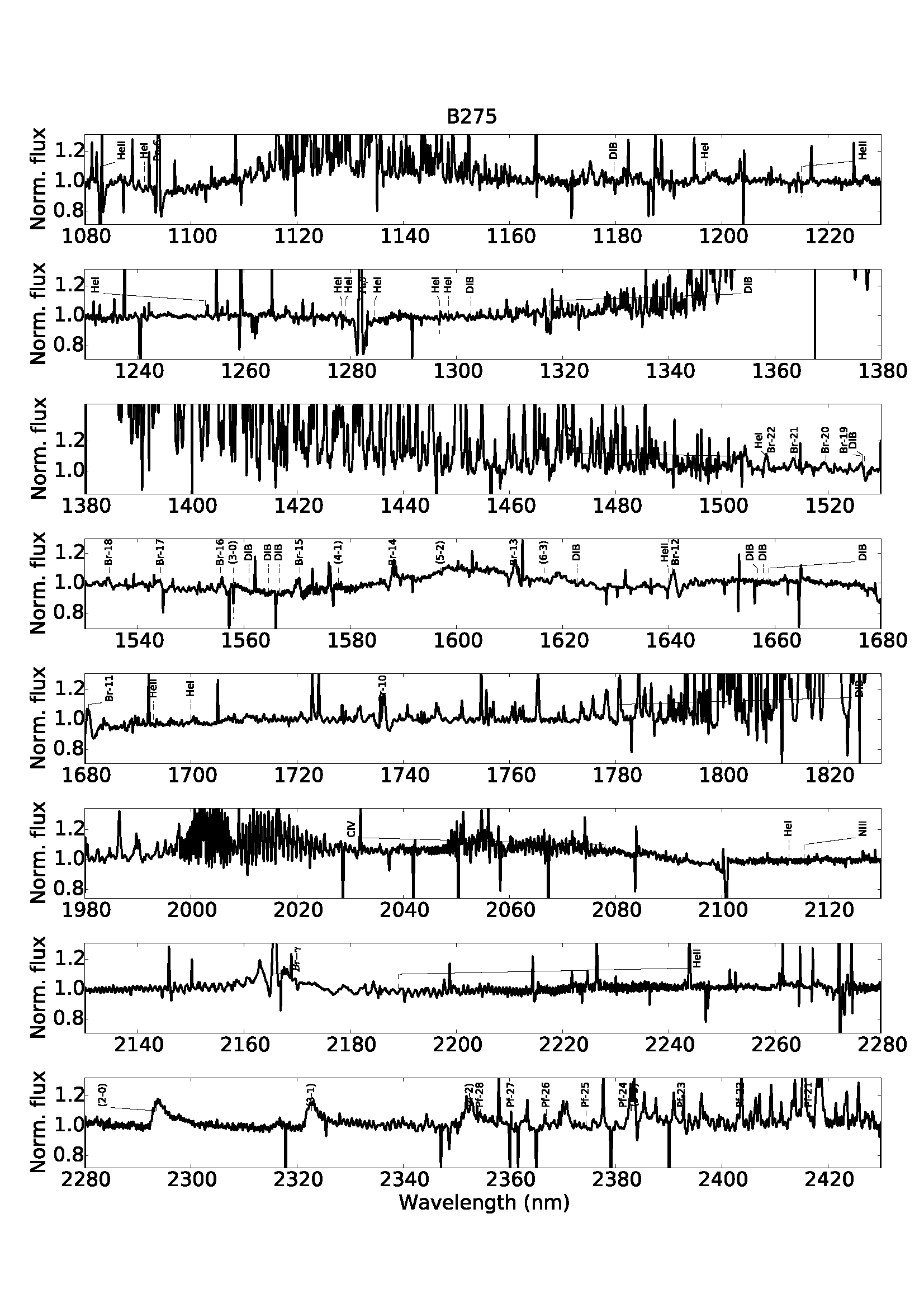}
   \end{figure}
%  \begin{figure}[h]
%  \includegraphics[width=\hsize]{figures/normspec_all_Page_23.png}
%   \end{figure}
%  \begin{figure}[h]
%  \includegraphics[width=\hsize]{figures/normspec_all_Page_24.png}
%   \end{figure}
%  \begin{figure}[h]
%  \includegraphics[width=\hsize]{figures/normspec_all_Page_25.png}
%   \end{figure}
%  \begin{figure}[h]
%  \includegraphics[width=\hsize]{figures/normspec_all_Page_26.png}
%   \end{figure}
  \begin{figure}[h]
  \includegraphics[width=\hsize]{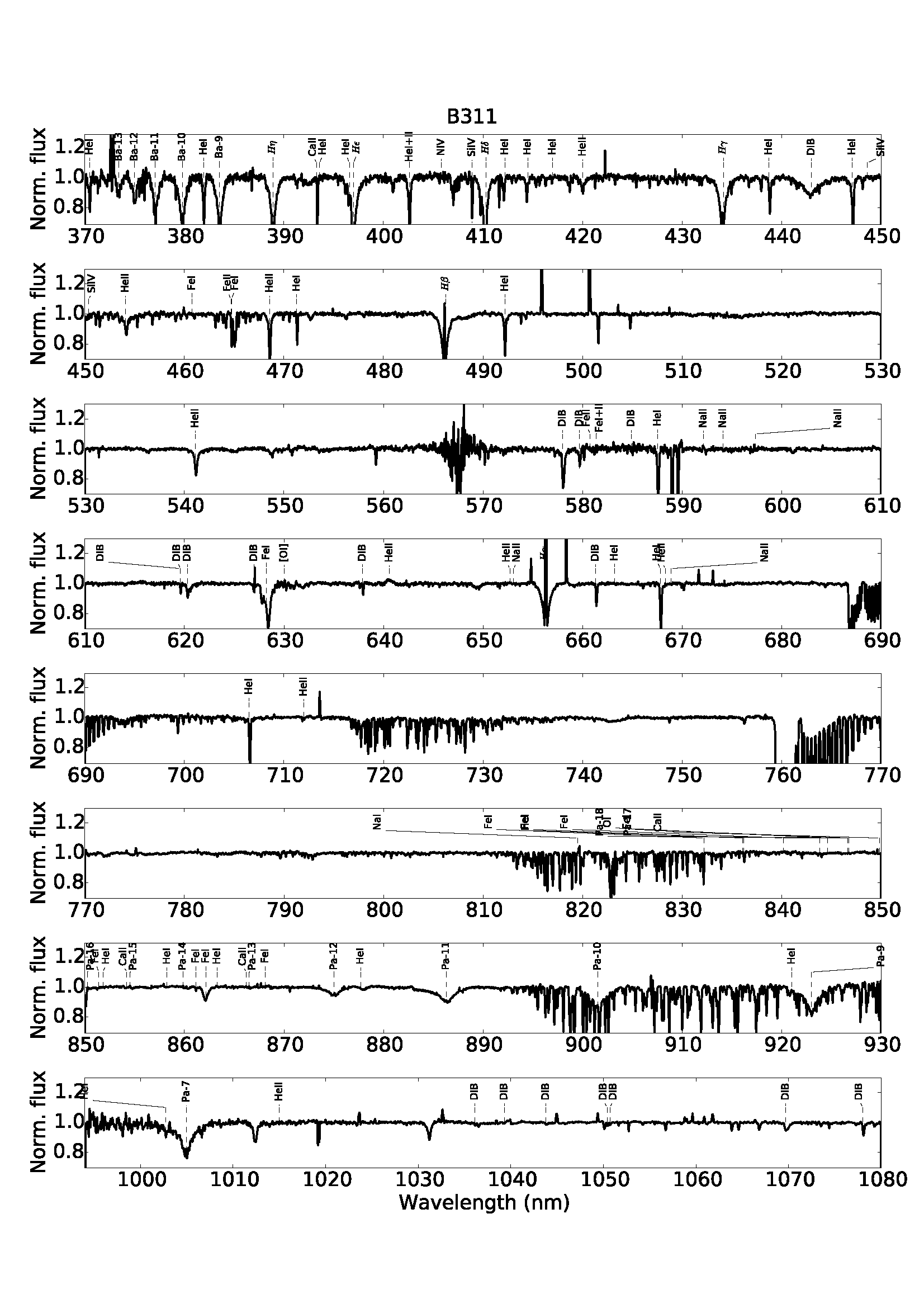}
   \end{figure}
  \begin{figure}[h]
  \includegraphics[width=\hsize]{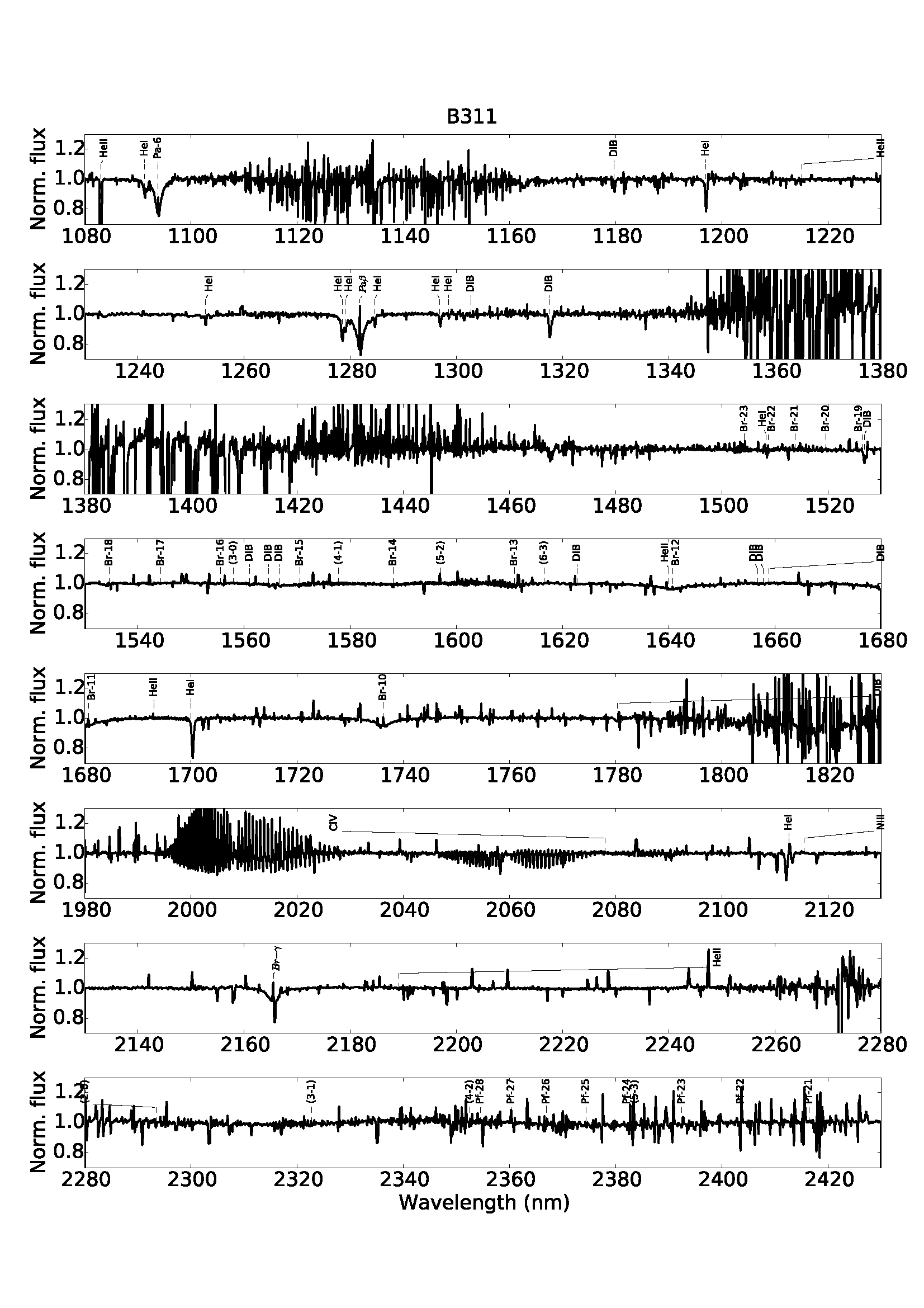}
   \end{figure}
%  \begin{figure}[h]
%  \includegraphics[width=\hsize]{figures/normspec_all_Page_29.png}
%   \end{figure}
%  \begin{figure}[h]
%  \includegraphics[width=\hsize]{figures/normspec_all_Page_30.png}
%   \end{figure}
%  \begin{figure}[h]
%  \includegraphics[width=\hsize]{figures/normspec_all_Page_31.png}
%   \end{figure}
%  \begin{figure}[h]
%  \includegraphics[width=\hsize]{figures/normspec_all_Page_32.png}
%   \end{figure}

%\includepdf[pages=-]{figures/normspec_all.pdf}

%------------------------------------------------
\clearpage
\newpage

\section{\texttt{FASTWIND} fitting results}
\label{P1:appendix_fastwind}

In this appendix we show in detail the model fits for our targets and the spectral lines used in each of the cases. We have included the models for B111 and B164 as an example, the version of the appendix for the full sample will be available on A\&A.

\vspace{-0.07\baselineskip}
\begin{figure*}[htb!]
     \centering
          \subfigure[B111]{%
            \label{P1:fig:FASTWIND111}
            \includegraphics[width=15.5cm]{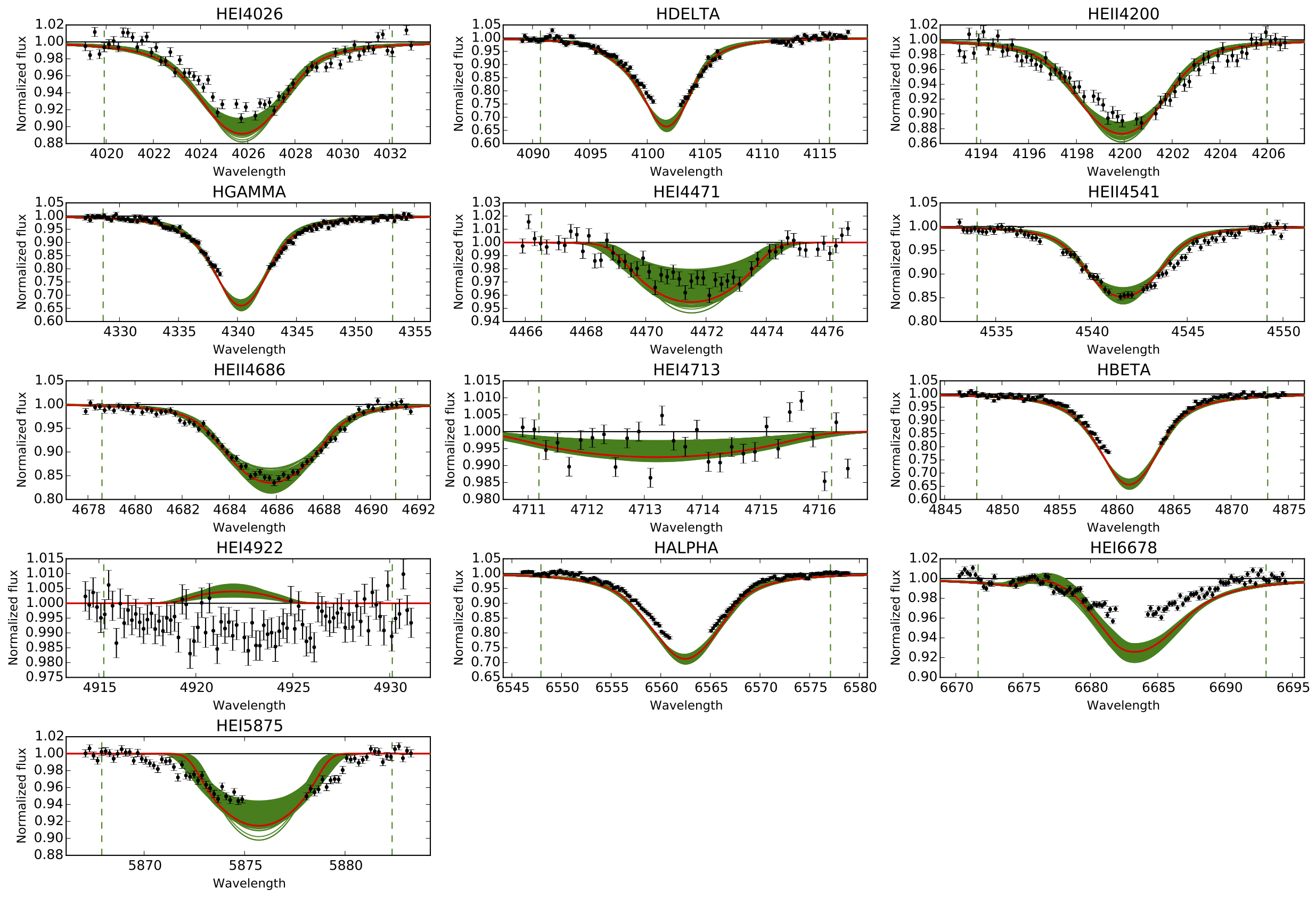}
        }\\%
        \vspace{-1.07\baselineskip}
          \subfigure[B164]{%
            \label{P1:fig:FASTWIND164}
            \includegraphics[width=15.5cm]{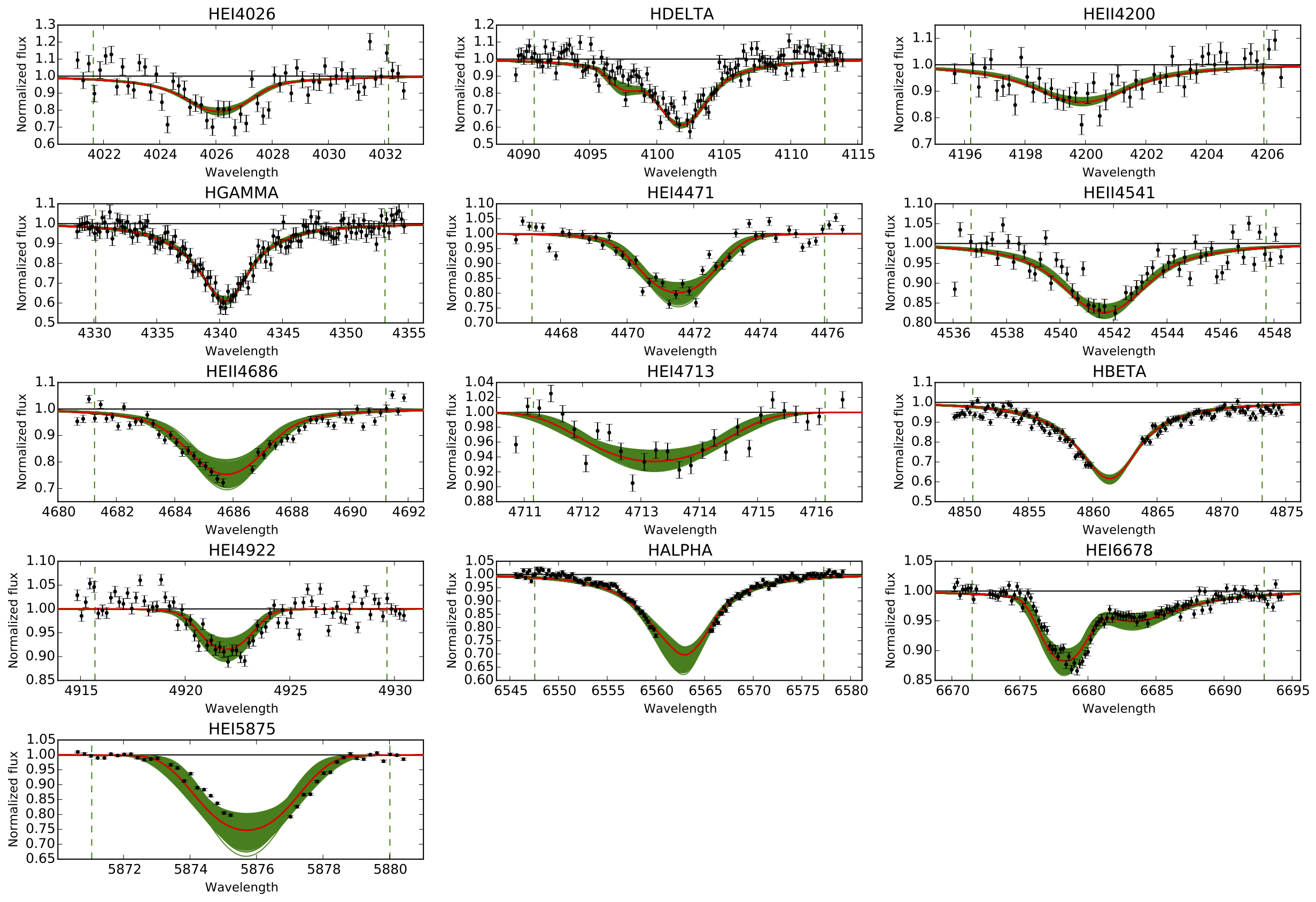}
        }%

    \caption[]{Observed spectra (black dots), the best fitting model (red), and other acceptable fits (5\% significance level or higher; green) for B111 and B164. The vertical dashed lines indicate the spectral range used for the fit and the horizontal black line gives the position of the continuum.}
\label{P1:fig:FASTWIND_fit_B111_B164}
\end{figure*}

\end{appendix}

\end{document}